\documentclass[preprint,9pt]{elsarticle}

\usepackage{amssymb}
\usepackage{amsmath}
\usepackage{float}
\usepackage{url}
\usepackage{xcolor}
\usepackage{bm}
\usepackage{algorithm}
\usepackage{algpseudocode}  
\usepackage{caption} 
\usepackage{graphicx}
\usepackage{mathtools}
\usepackage{booktabs}
\usepackage{doi}
\usepackage{listings}
\usepackage{multirow}
\usepackage{plantuml}
\usepackage{subcaption}
\usepackage{makecell}

\journal{ACM Transaction on Mathematical Software}

\begin{document}

\begin{frontmatter}

\title{OpenLB-UQ: An Uncertainty Quantification Framework for Incompressible Fluid Flow Simulations}

\author[SCC,LBRG,IANM]{Mingliang Zhong}
\author[LBRG,IANM]{Adrian Kummerl\"{a}nder}
\author[LBRG,IANM]{Shota Ito}
\author[LBRG,IANM,MVM]{Mathias J. Krause}
\author[SCC,IANM]{Martin Frank}
\author[LBRG,IANM]{Stephan Simonis}

\affiliation[SCC]{organization={Scientific Computing Center, Karlsruhe Institute of Technology},
            addressline={Eggenstein-Leopoldshafen}, 
            city={Karlsruhe},
            postcode={76344}, 
            state={Baden-Württemberg},
            country={Germany}}
            
\affiliation[IANM]{organization={Institute for Applied and Numerical Mathematics, Karlsruhe Institute of Technology},
            addressline={Englerstr. 2}, 
            city={Karlsruhe},
            postcode={76131}, 
            state={Baden-Württemberg},
            country={Germany}}

\affiliation[MVM]{organization={Institute of Mechanical Process Engineering and Mechanics},
            addressline={Straße am Forum 8}, 
            city={Karlsruhe},
            postcode={76131}, 
            state={Baden-Württemberg},
            country={Germany}}

\affiliation[LBRG]{organization={Lattice Boltzmann Research Group, Karlsruhe Institute of Technology},
            addressline={Englerstr. 2}, 
            city={Karlsruhe},
            postcode={76131}, 
            state={Baden-Württemberg},
            country={Germany}}            

\begin{abstract}
Uncertainty quantification (UQ) is crucial in computational fluid dynamics to assess the reliability and robustness of simulations, given the uncertainties in input parameters. 
OpenLB is an open-source lattice Boltzmann method library designed for efficient and extensible simulations of complex fluid dynamics on high-performance computers. 
In this work, we leverage the efficiency of OpenLB for large-scale flow sampling with a dedicated and integrated UQ module. 
To this end, we focus on non-intrusive stochastic collocation methods based on generalized polynomial chaos and Monte Carlo sampling. 
The OpenLB-UQ framework is extensively validated in convergence tests with respect to statistical metrics and sample efficiency using selected benchmark cases, including two-dimensional Taylor--Green vortex flows with up to four-dimensional uncertainty and a flow past a cylinder. 
Our results confirm the expected convergence rates and show promising scalability, demonstrating robust statistical accuracy as well as computational efficiency. 
OpenLB-UQ enhances the capability of the OpenLB library, offering researchers a scalable framework for UQ in incompressible fluid flow simulations and beyond.
\end{abstract}

\begin{highlights}
\item OpenLB-UQ: first open-source lattice Boltzmann method framework with built-in non-intrusive uncertainty quantification.
\item Stochastic collocation generalized polynomial chaos achieves spectral convergence on flow around a cylinder and Taylor--Green vortex benchmarks.
\item Hybrid sample- and domain-level parallelism scales efficiently, reducing UQ wall-clock time by a factor $>30$.
\end{highlights}


\end{frontmatter}

\tableofcontents

\section{Introduction}
\label{}

Uncertainty quantification (UQ) is crucial in computational fluid dynamics (CFD), as it enables the assessment of the reliability and robustness of simulation results. 
CFD simulations are inherently affected by uncertainties arising from imprecise boundary conditions, variable operating environments, and uncertain material properties—factors that are rarely known with complete accuracy~\cite{najm2009uncertainty}. 
UQ methods provide the statistical foundation to propagate these uncertainties through simulations, transforming CFD from a purely predictive tool into a robust instrument for decision-making. 
Accurate treatment of uncertainty is fundamental for defining design margins, optimizing control strategies, and conducting effective risk assessments. 
Despite its importance and the increasingly efficient code developments in the last decades, open-source CFD frameworks that offer integrated support for advanced UQ techniques remain scarce. 
This paper fills that gap by presenting OpenLB‑UQ, the first fully integrated, open‑source lattice Boltzmann library that offers native support for state‑of‑the‑art non‑intrusive UQ techniques. 

Among the various branches of CFD, incompressible fluid flow simulations are widely used in engineering applications, including automotive, aerospace, and biomedical engineering. 
Incompressible flows characterized by constant fluid density and divergence-free velocity fields introduce unique challenges for UQ due to their pronounced sensitivity to input uncertainties. 
This sensitivity can significantly influence critical quantities of interest, such as velocity and pressure fields~\cite{xiu2003modeling,mccullough2024uncertainty}, as well as derived aerodynamic forces like drag and lift coefficients~\cite{oliver2011bayesian}.

Common sources of uncertainty in incompressible flow simulations include variations in geometric tolerances~\cite{marepally2022uncertainty}, inflow boundary conditions~\cite{yu2022inlet}, and material properties such as viscosity and density.
Accurately quantifying and propagating these uncertainties is essential for designing robust systems and for interpreting simulation results with greater confidence.

Traditional CFD approaches typically involve the numerical solution of the governing equations of fluid flow most notably, the Navier--Stokes equations (NSE) to predict fluid behavior.
While these methods are widely adopted, they can be computationally intensive and difficult to scale efficiently on high-performance computing platforms, particularly when UQ is incorporated.

The lattice Boltzmann method (LBM) offers a mesoscopic alternative to traditional CFD techniques by simulating fluid behavior through particle distribution functions rather than directly solving the NSE~\cite{succi2001lattice,lallemand2021lattice,guo2013lattice,junk2005asymptotic}.
LBM has gained considerable popularity due to its algorithmic simplicity, inherent parallelism, and effectiveness in handling complex geometries and boundary conditions.
OpenLB, a continuously developed open-source LBM framework, provides comprehensive support for parallel computing through MPI, OpenMP, and CUDA, enabling efficient execution on a wide range of hardware from desktop workstations to supercomputers~\cite{olbPaper2021,olbRelease18}. 
Its modular architecture and extensive model library have enabled diverse applications, including multiphase, turbulent~\cite{simonis2022temporal,molinaro2025generative}, thermal~\cite{ross2021pore}, porous media~\cite{ross2019conjugate}, and particle-laden flows~\cite{hafen2022simulation,bukreev4991224prediction}.
These capabilities make OpenLB a robust platform for developing and validating novel methodologies—such as the UQ extension framework presented in this work.

Monte Carlo sampling (MCS), often considered the benchmark for UQ, requires a large number of simulation evaluations due to its relatively slow convergence rate of $\mathcal{O}(N^{-0.5})$, making its direct application computationally expensive for complex CFD problems~\cite{lucor2007stochastic}.
To partially mitigate this issue, quasi Monte Carlo (QMC) method addresses this limitation by using low-discrepancy, deterministic sampling points~\cite{morokoff1995quasi}.
Nevertheless, the convergence rate of QMC methods remains limited to approximately $\mathcal{O}(N^{-1})$, which is still not sufficiently fast for many practical engineering scenarios. 

Consequently, alternative UQ approaches capable of significantly accelerating convergence are necessary. Generalized polynomial chaos (gPC) methods have emerged as attractive candidates because of their spectral convergence rates for smooth problems, dramatically reducing the required number of simulation evaluations.
The gPC approach can be categorized into intrusive methods, such as stochastic Galerkin (SG-gPC), and non-intrusive methods, such as stochastic collocation (SC-gPC)~\cite{mathelin2005stochastic,xiu2005high,babuvska2007stochastic}.
While SG-gPC achieves highly accurate solutions, it requires extensive modifications throughout the underlying CFD solver, making its implementation challenging or infeasible within established simulation frameworks~\cite{zhong2024stochastic}. 

In contrast, SC-gPC is a non-intrusive, sampling-based method that requires minimal changes to existing solvers.
Building on this advantage, we present a dedicated UQ extension for OpenLB that integrates SC-gPC along with other non-intrusive techniques.
This extension enables efficient and scalable UQ within a high-performance lattice Boltzmann framework.
By reusing a carefully optimized solver for each sample and minimizing additional overhead, our approach maintains solver modularity while supporting extensive UQ analysis. 

To evaluate the capabilities of the proposed OpenLB-UQ extension, we apply it to a series of representative test cases that demonstrate both fundamental correctness and practical applicability. The 2D Taylor--Green vortex, a canonical benchmark for incompressible flow solvers, serves to validate the implementation against its analytical solution. A variant of this case, incorporating four-dimensional uncertainty, further tests the framework’s ability to handle high-dimensional stochastic initial conditions. Additionally, the flow past a cylinder scenario evaluates the extension’s effectiveness in capturing uncertainty propagation in aerodynamic quantities such as drag coefficients and wake structures under stochastic inflow conditions. Collectively, these cases illustrate the robustness, accuracy, and versatility of the OpenLB-UQ extension in addressing a variety of UQ challenges. 
In addition, the framework can readily be used as a UQ-wrapper for any other application case in OpenLB, which includes a variety of complex fluid flow problems as well as multiphysics applications, both stationary and time-dependent in up to three spatial dimensions. 

Results across these test cases consistently validate the computational effectiveness and reliability of the SC-gPC based UQ framework integrated into OpenLB.
By combining high computational efficiency with ease of integration, OpenLB-UQ provides a practical, fully open-source platform, eliminating the need for external UQ libraries and lowering the barrier for adoption in the CFD research community.

In conclusion, the integration of non-intrusive UQ methods into OpenLB establishes a unified, efficient, and extensible solution for propagating uncertainties in incompressible flow simulations.
The OpenLB-UQ extension enhances both reliability and performance while preserving the flexibility and modularity of the existing framework.
Future work will focus on expanding support for complex flow scenarios and incorporating hybrid UQ strategies to further improve accuracy and scalability.

This paper is structured as follows: 
Section~\ref{sec:UINSE} summarizes the uncertain problem formulation for incompressible fluid flow. 
Section~\ref{sec:lbm} recalls the underlying deterministic LBM. 
Section~\ref{sec:uq} gives an overview of the UQ methods used in our framework.
Section~\ref{sec:softwareArch} summarizes the software architecture of OpenLB-UQ and Section~\ref{sec:core_class_design} provides a detailed documentation of the core classes. 
In Section~\ref{sec:runtime_workflow} we sketch the runtime workflow. 
Finally, Section~\ref{sec:results} presents the numerical experiments and an error-based validation, Section~\ref{sec:perf} includes a detailed performance evaluation and Section~\ref{sec:conclusion} draws conclusions and suggests future work. 

\section{Incompressible Navier--Stokes equations with uncertainty} \label{sec:UINSE}
To systematically account for uncertainty in fluid dynamics simulations, we begin with the incompressible NSE
\begin{align}
    \partial_t \bm{u} + (\bm{u} \cdot \bm{\nabla}) \bm{u} - \nu \bm{\nabla}^2 \bm{u} = -\frac{1}{\rho} \bm{\nabla} p + \bm{g}, 
        \quad & \text{in } \mathcal{X} \times \mathcal{I} \subseteq \mathbb{R}^{d_{x}} \times \mathbb{R}_{> 0}, \label{eq:detNSEmom} \\
    \bm{\nabla} \cdot \bm{u} = 0, 
        \quad &\text{in } \mathcal{X} \times \mathcal{I}, \label{eq:detNSEdiv}
\end{align}
with appropriate boundary and initial conditions, which model the behavior of Newtonian fluids at low Mach numbers. 
Here, $\rho > 0$ denotes fluid density, $\bm{u}\colon \mathcal{X}\times \mathcal{I} \to \mathcal{U} \subseteq \mathbb{R}^{d_{x}}$ represents the velocity vector, $p\colon \mathcal{X} \times \mathcal{I} \to \mathbb{R}$ is the pressure, $\nu>0$ is the kinematic viscosity, and $\bm{g}\colon \mathcal{X}\times \mathcal{I} \to \mathbb{R}^{d_{x}} $ represents an external force. 
In our study, the external force is ignored (\(\bm{g} = \bm{0}\)).

To incorporate uncertainty, we extend the deterministic NSE~\eqref{eq:detNSEmom},~\eqref{eq:detNSEdiv} to a stochastic framework by introducing random variables into selected input parameters.

Let $\bm{Z} = (Z_1, ..., Z_{d_{Z}}) \in \mathcal{Z} \subseteq \mathbb{R}^{d_Z}$  be a vector of independent random variables defined on a probability space \((\Omega, \mathcal{F}, \mathbb{P})\). 
The stochastic form of the NSE thus reads
\begin{align}
    \partial_t \bm{u}(\bm{Z}) + (\bm{u}(\bm{Z})\cdot \bm{\nabla}) \bm{u}(\bm{Z}) - \nu(\bm{Z}) \bm{\nabla}^2 \bm{u}(\bm{Z}) = -\frac{1}{\rho} \bm{\nabla} p(\bm{Z}), \quad & \text{in } \mathcal{X} \times \mathcal{I} \times \mathcal{Z}, \label{eq:stochNSEmom} \\
    \bm{\nabla} \cdot \bm{u}(\bm{Z}) = 0, \quad & \text{in } \mathcal{X} \times \mathcal{I} \times \mathcal{Z}, \label{eq:stochNSEdiv}
\end{align}
where we neglected the space-time arguments \((\bm{x},t)\in \mathcal{X} \times \mathcal{I}\) of \(\bm{u}\) for the sake of readability.
In this work, we consider both one- and multi-dimensional uncertainties. 
Two representative cases are examined, based on uncertainty in the velocity field $\bm{u}(\bm{Z})$ induced by (i) stochastic boundary or initial conditions or (ii) uncertainty in the viscosity $\nu(\bm{Z})$. 

Notably, for a fixed vector \(\bm{Z}\), the stochastic NSE~\eqref{eq:stochNSEmom},~\eqref{eq:stochNSEdiv} becomes a version of the deterministic NSE~\eqref{eq:detNSEmom},~\eqref{eq:detNSEdiv}, which forms the basis of the non-intrusive methodology in our framework.

\section{Lattice Boltzmann methods} \label{sec:lbm}

We use the LBM to numerically solve the deterministic NSE~\eqref{eq:detNSEmom},~\eqref{eq:detNSEdiv}. 
The core of the deterministic LBM is the lattice Boltzmann equation (LBE), given by
\begin{align}\label{eq:deterministicLBE}
f_{i}\left( {\bm{x} + \bm{c}_{i}\triangle t,t + \triangle t} \right) - f_{i}\left( {\bm{x},t} \right) = \Omega_{i}(\bm{f}(\bm{x}, t)) , \quad \text{in } \mathcal{X}_{\triangle x}\times \mathcal{I}_{\triangle t} ,
\end{align}
where \(i=0, 1, \ldots, q-1\), the population vector $\bm{f} \in \mathbb{R}^{q}$ represents the discretized particle distribution function, $\bm{c}_i\in \mathbb{R}^{d_{x}}$ denotes the particle speed in the $i$th velocity direction, and $\bm{\Omega}(\bm{f})\in \mathbb{R}^{q}$ represents the collision term.

For a review of the LBM for fluid dynamics problems, we refer the reader to Lallemand~\textit{et al.}~\cite{lallemand2021lattice}. 
In general, the LBE~\eqref{eq:deterministicLBE} can be derived, for example, from a finite difference discretization of a discrete velocity Boltzmann equation on a uniform space-time grid with discretization steps \(\triangle x\) and \(\triangle t\) (see e.g.\ \cite{simonis2022limit} and references therein). 
Here, we use a Bhatnagar--Gross--Krook (BGK)~\cite{bhatnagar1954model} collision model
\begin{align} \label{eq:bgkCollision}
    \Omega_{i}(\bm{f}(\bm{x}, t)) = - \frac{1}{\tau} \left( f_{i}\left( {\bm{x},t} \right) - f_{i}^{\mathrm{eq}}\left( {\bm{x},t} \right) \right) .
\end{align}
In \eqref{eq:bgkCollision}, $\tau>0.5$ denotes the relaxation time, while $f_{i}^{\mathrm{eq}}$ represents the equilibrium distribution function given by 
\begin{align} \label{equilibrium}
f_{i}^{\mathrm{eq}}\left( \bm{x},t \right) = w_{i}\rho\left( \bm{x},t \right)\left( {1 + \frac{\bm{u}\left( \bm{x},t \right)\cdot \bm{c}_{i}}{c_{s}^{2}} + \frac{\left( {\bm{u} \left( \bm{x},t \right)\cdot \bm{c}_{i}} \right)^{2}}{2c_{s}^{4}} - \frac{\bm{u}\left( \bm{x},t \right) \cdot \bm{u}\left( \bm{x},t \right)}{2c_{s}^{2}}} \right) ,
\end{align}
where $w_i$ is the weight of each discrete velocity and $c_s$ represents the sound speed.

The density $\rho$ and the momentum density $\rho \bm{u}$ are calculated as moments of $f_i$ via 
\begin{align}
\rho\left( \bm{x},t \right) &= \sum\limits_{i=0}^{q-1} f_{i}\left( \bm{x},t \right), \\
\rho\left( \bm{x},t \right) \bm{u}\left( \bm{x},t \right) &= \sum\limits_{i=0}^{q-1}\bm{c}_{i} f_{i} \left( \bm{x},t \right) , 
\end{align}
and (in our version of the scheme) approximate the macroscopic unknowns in \eqref{eq:detNSEmom},~\eqref{eq:detNSEdiv} up to second and first order in space, respectively~\cite{simonis2022limit}.

The discrete velocity models used in the LBM are denoted as \(Dd_{x}Qq\) and characterized by the spatial dimension $d_{x}$ and the number of discrete velocities $q$.

The present study is based on the \(D2Q9\) velocity discrete model, where the nine lattice velocities are expressed as
\begin{align}
\bm{c}_{i} 
= 
\begin{cases} 
 (0,0), & \quad \text{if } i=0, \\
 (\pm 1,0), (0, \pm1), & \quad \text{if } i=1, 2, \ldots, 4, \\
 (\pm 1, \pm 1), & \quad \text{if } i=5,6, \ldots, 8 , 
\end{cases} 
\end{align}
with corresponding weights 
\begin{align}\label{eq:weightsDdQ9}
w_i 
= 
w \begin{cases} 
    	\frac{4}{9},  & \quad \text{if } i=0, \\
    	\frac{1}{9},  & \quad \text{if } i=1,2,\ldots,4, \\
    	\frac{1}{36}, & \quad \text{if } i=5,6,\ldots,8 ,
	\end{cases}
\end{align}
where \(w\) is the classical weight function in LBM \cite{lallemand2021lattice}. 
The LBE \eqref{eq:deterministicLBE} yields a perfectly parallelizable numerical scheme, when decomposed into two distinct steps, reading
\begin{align}    
\text{collision:} & \quad f_{i}^{\star}\left( \bm{x},t \right) = f_{i}\left( \bm{x},t \right) - \frac{1}{\tau}\left( f_{i}\left( \bm{x},t \right) - f_{i}^{\mathrm{eq}}\left( {\bm{x},t} \right) \right),  \\
\text{streaming:} & \quad f_{i}\left( \bm{x} + \bm{c}_{i}\triangle t,t + \triangle t \right) = f_{i}^{\star}\left( \bm{x},t \right) ,
\end{align}
respectively, where the upper index \(\cdot^{\star}\) denotes post-collision variables.

Dependent on the initial and boundary conditions for the target problem \eqref{eq:detNSEmom}, \eqref{eq:detNSEdiv}, suitable LBM-specific initialization and boundary methods for \eqref{eq:deterministicLBE} can be used \cite{simonis2023lattice}.

In summary, from a kinetic viewpoint, the LBM discretizes velocity, space, and time of the continuous BGK Boltzmann equation. 
The space-time advancement is achieved through the successive evolution of collision and streaming processes, allowing the simulation of fluid flow phenomena and beyond with high computational efficiency in optimized implementations such as in OpenLB~\cite{olbPaper2021,kummerlander2022advances}.

\section{Uncertainty quantification methods}\label{sec:uq}

In this section, we summarize the UQ methodology used in the OpenLB-UQ framework. 
The UQ methods are then combined with the deterministic LBM given in Section~\ref{sec:lbm} to numerically solve problems that are modeled e.g.\ by \eqref{eq:stochNSEmom},~\ref{eq:stochNSEdiv} with additional boundary and initial conditions. 

\subsection{Monte Carlo and quasi Monte Carlo sampling}

The MCS method is the most straightforward and widely accepted benchmark for UQ.
It estimates statistical moments by randomly sampling the input parameters and repeatedly evaluating the numerical solver. 

Let \( \bm{Z} \in \mathbb{R}^{d_Z} \) denote a random vector representing the uncertain inputs, and suppose \( \mathcal{M}(\bm{Z}) \in \mathbb{R} \) denotes a scalar model response or quantity of interest (e.g., a velocity component or drag coefficient) that depends on \( \bm{Z} \).

Given a set of \( N_q \) samples \( \{\bm{Z}^{(n)}\}_{n=1}^{N_q} \), drawn either randomly (MCS) or deterministically using low-discrepancy sequences (QMC), the first- and second-order statistical moments of \( \mathcal{M} \) are approximated by
\begin{align}
\mathbb{E}[\mathcal{M}(\bm{Z})] & \approx \frac{1}{N_q} \sum_{n=1}^{N_q} \mathcal{M}(\bm{Z}^{(n)}), \label{eq:meanMCS} \\
\text{Var}[\mathcal{M}(\bm{Z})] & \approx \frac{1}{N_q-1} \sum_{n=1}^{N_q} \left( \mathcal{M}(\bm{Z}^{(n)}) - \frac{1}{N_q} \sum_{m=1}^{N_q} \mathcal{M}(\bm{Z}^{(m)}) \right)^2. \label{eq:varMCS}
\end{align}

In the QMC method, the same estimators \eqref{eq:meanMCS} and \eqref{eq:varMCS} are used, but the samples \( \{\bm{Z}^{(n)}\}_{n=1}^{N_q} \) are generated using low-discrepancy sequences such as Sobol or Halton. These sequences more uniformly cover the input space than random samples, often yielding faster convergence for smooth integrands and moderate stochastic dimensions.

MCS converges at a dimension-independent rate of \( \mathcal{O}({N_q}^{-1/2}) \), which is often too slow for high-fidelity CFD simulations. 
In contrast, QMC achieves a rate of \( \mathcal{O}\left( (\log {N_q})^{d_Z} {N_q}^{-1} \right) \), offering superior performance in lower dimensions~\cite{caflisch1998monte}.

However, the exponential dependence on \( d_Z \) may hinder efficiency in high-dimensional problems.

Although this improved rate suggests superior performance in lower dimensions, the exponential growth of the logarithmic term can degrade QMC efficiency in high-dimensional settings.

Nevertheless, by applying dimension-reduction strategies or appropriate transformations to the integrand, QMC often outperforms MCS in practical applications—both in accuracy and computational cost.

In this work, we adopt MCS as a reference configuration and as a fallback method when other sampling strategies fail or become impractical.

\subsection{Generalized polynomial chaos}

As efficient alternatives to MCS, gPC methods offer a more computationally tractable approach for UQ with high accuracy. The gPC framework is particularly well-suited for problems involving smooth dependence of the model response on a moderate number of uncertain parameters.

Let \( \bm{Z} \in \mathbb{R}^{d_Z} \) be a random vector with joint probability density function \( h(\bm{Z}) \), where \( d_Z \) is the stochastic dimension.
The gPC method approximates \( \mathcal{M}(\bm{Z}) \) using a finite expansion of orthonormal polynomial basis functions \( \{\Phi_k(\bm{Z})\} \), tailored to the distribution of \( \bm{Z} \), as follows:
\begin{equation} \label{eq:gPC}
    \mathcal{M}(\bm{Z}) \approx \mathcal{M}^{(N)}(\bm{Z}) 
    = \sum_{k=0}^{P} \hat{\mathcal{M}}_{k}\,\Phi_{k}(\bm{Z}),
\end{equation}
where \( P \) is the highest polynomial index in the truncated expansion, and \( \hat{g}_{k} \) are the associated expansion coefficients.

These coefficients are computed via the projection of \( \mathcal{M}(\bm{Z}) \) onto the basis functions, i.e.
\begin{equation} \label{eq:coef_integral}
    \hat{\mathcal{M}}_{k} 
    = \frac{1}{\gamma_k} \mathbb{E}\left[ \mathcal{M}(\bm{Z})\,\Phi_k(\bm{Z}) \right]
    = \frac{1}{\gamma_k} \int_{\mathbb{R}^{d_Z}} \mathcal{M}(\bm{Z})\,\Phi_k(\bm{Z})\,h(\bm{Z})\,\mathrm{d}\bm{Z},
\end{equation}
where \( \gamma_k = \mathbb{E}[\Phi_k^2(\bm{Z})] \) is the normalization constant (equal to 1 for orthonormal bases).

In practice, the integral in Eq.\eqref{eq:coef_integral} is approximated using a quadrature rule. Let \( \{\bm{Z}^{(j)}\}_{j=1}^{{N_q}} \subseteq \mathbb{R}^{d_Z} \) denote the quadrature (or collocation) nodes, and \( \{a_j\}_{j=1}^{N_q} \) the associated scalar weights. The expansion coefficients are then approximated by
\begin{equation} \label{eq:coef_discrete}
    \hat{\mathcal{M}}_k 
    \approx \frac{1}{\gamma_k} \sum_{j=1}^{{N_q}} a_j\,\mathcal{M}(\bm{Z}^{(j)})\,\Phi_k(\bm{Z}^{(j)}),
\end{equation}
where \( N \) is the total number of quadrature points used in the stochastic space.

Based on the preceding formulation, the SC-gPC method constructs a surrogate approximation of the model response \( \mathcal{M}(\bm{Z}) \) by evaluating it at specific realizations (collocation points) of the random vector \( \bm{Z} \). The procedure involves the following steps:
\begin{enumerate}
    \item Determine the set of collocation points \( \{\bm{Z}^{(j)}\}_{j=1}^{{N_q}} \subset \mathbb{R}^{d_Z} \) and corresponding weights \( \{a_j\}_{j=1}^{{N_q}} \) based on a suitable quadrature rule adapted to the distribution \( h(\bm{Z}) \).
    \item At each collocation point \( \bm{Z}^{(j)} \), solve the deterministic model to obtain the response \( \mathcal{M}(\bm{Z}^{(j)}) \).
    \item Approximate the gPC expansion coefficients \( \hat{\mathcal{M}}_k \) using the discrete projection formula in Eq.~\eqref{eq:coef_discrete}.
    \item Construct the polynomial surrogate \( \mathcal{M}^{(N)}(\bm{Z}) \) using the truncated gPC expansion in Eq.~\eqref{eq:gPC}.
\end{enumerate}

Once the coefficients \( \{ \hat{\mathcal{M}}_k \} \) are computed, statistical moments can be directly extracted due to the orthonormality of the basis \( \{ \Phi_k \} \) with respect to \( h(\bm{Z}) \). The mean of \( \mathcal{M}(\bm{Z}) \) is given by
\begin{align} \label{eq:gpc_mean}
    \mathbb{E}[\mathcal{M}(\bm{Z})] 
    = \int_{\mathbb{R}^{d_Z}} \mathcal{M}(\bm{Z}) h(\bm{Z})\, \mathrm{d}\bm{Z} 
    \approx \hat{\mathcal{M}}_{0},
\end{align}
since \( \Phi_0(\bm{Z}) = 1 \) and all other basis functions are orthogonal to it.

The variance is approximated by summing the squared coefficients of the higher-order modes:
\begin{align}
    \text{Var}[\mathcal{M}(\bm{Z})] 
    = \mathbb{E}[\mathcal{M}^2(\bm{Z})] - \mathbb{E}[\mathcal{M}(\bm{Z})]^2 
    \approx \sum_{k=1}^{N} \hat{\mathcal{M}}_k^2.
\end{align}
Accordingly, the standard deviation is estimated as
\begin{equation} \label{eq:gpc_std}
    \sigma(\mathcal{M}) \approx \left( \sum_{k=1}^{N} \hat{\mathcal{M}}_k^2 \right)^{1/2}.
\end{equation}

These expressions follow directly from the orthonormality of the polynomial basis, which simplifies the evaluation of moments to operations on the expansion coefficients.

In summary, the SC-gPC method provides a non-intrusive approach to constructing polynomial surrogates for stochastic problems.
It avoids the need to modify the underlying deterministic solver (in contrast to intrusive SG methods), while retaining spectral convergence properties: the accuracy of the surrogate improves exponentially with increasing polynomial order \( P \), provided the model response \( \mathcal{M}(\bm{Z}) \) is sufficiently smooth in the stochastic space.

\section{Software architecture}\label{sec:softwareArch}

\begin{figure}[ht!]
    \centerline{
    \includegraphics[width=1.5\linewidth]{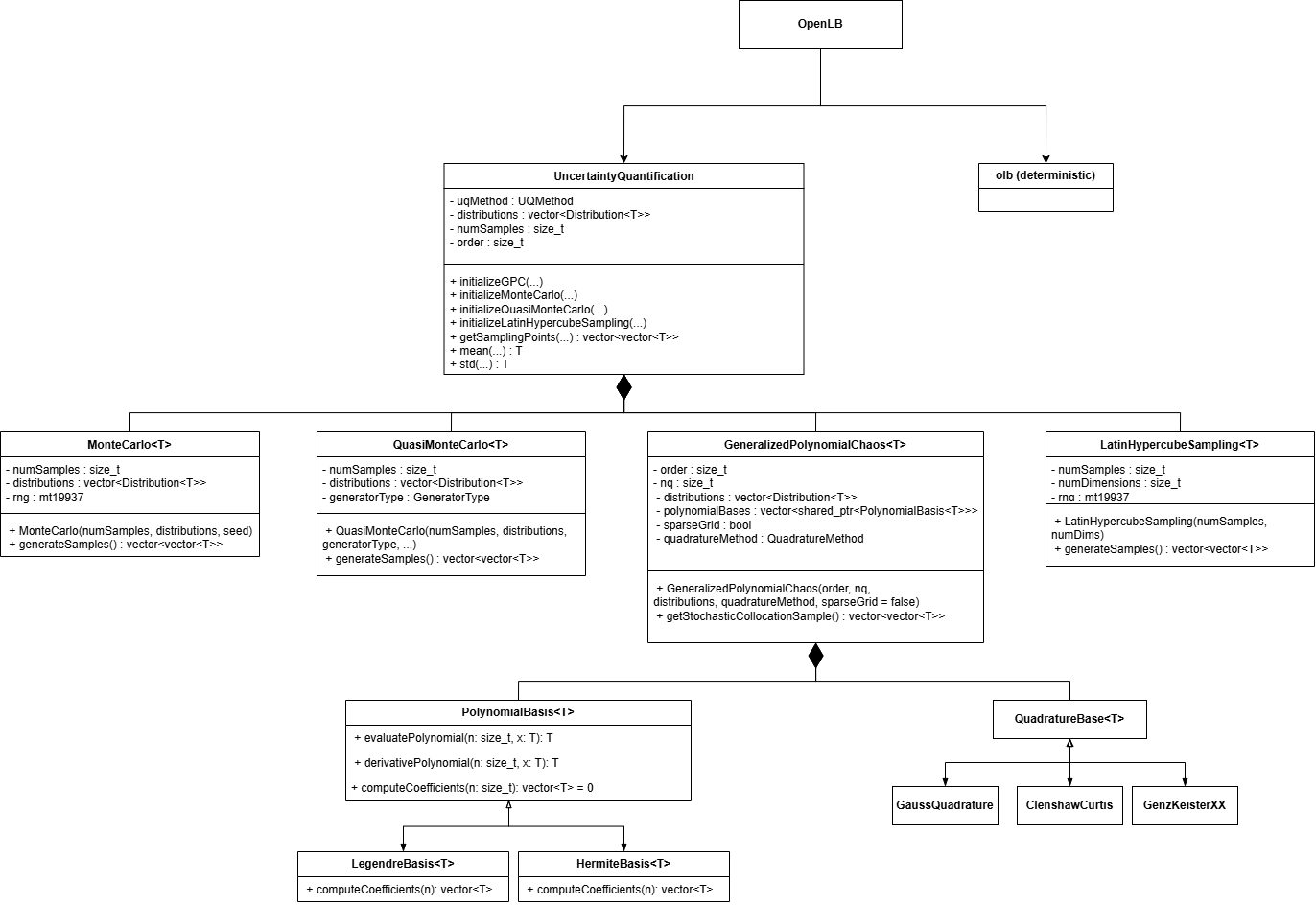}
    }
    \caption{Class hierarchy for OpenLB-UQ framework.}
    \label{fig:uml}
\end{figure}
The software architecture of the proposed framework is designed to enable UQ on top of the existing deterministic LBM simulations provided by OpenLB. 
Instead of modifying the internal solver kernel, the UQ layer operates as a high-level orchestrator that coordinates the sampling process, invokes the existing LBM application for each realization, and subsequently aggregates statistical results.

As illustrated in Figure~\ref{fig:uml}, the architecture consists of two loosely coupled layers:
\begin{itemize}
    \item \textbf{UQ layer:} 
    This layer is responsible for generating samples from uncertain inputs, managing the configuration of UQ methods, and performing statistical postprocessing. 
    It includes components such as \texttt{Distribution}, \texttt{SamplingStrategy}, and \texttt{QuadratureRule}, all orchestrated by the manager class called \texttt{UncertaintyQuantification}. 
    The UQ layer is agnostic to the underlying LBM discretization and solver details.
    \item \textbf{OpenLB layer:} 
    This layer runs the deterministic simulation for each sample provided by the UQ layer. 
    It utilizes OpenLB’s existing modular structure, including \texttt{SuperLattice}, \texttt{BlockLattice}, and user-defined boundary/initial conditions. 
    From the perspective of the LBM solver, each realization is simply a deterministic run with a specific set of input parameters.
\end{itemize}

Based on the above architecture, the overall execution follows a three-phase workflow:
\begin{enumerate}
    \item \textbf{Sampling:} 
    The UQ module generates a set of \( {N_q} \) samples \( \{ \bm{Z}^{(i)} \}_{i=1}^{{N_q}} \subset \mathbb{R}^{d_Z} \) from the prescribed input distribution using a chosen sampling or quadrature method (e.g., Monte Carlo, quasi-Monte Carlo, or gauss quadrature).
    
    \item \textbf{Simulation:} 
    Each sample \( \bm{Z}^{(i)} \) is passed to the OpenLB application, which performs a deterministic simulation and returns a scalar quantity of interest \( \mathcal{M}(\bm{Z}^{(i)}) \), such as the drag coefficient or probe velocity.

    \item \textbf{Postprocessing:} 
    The UQ module collects the model outputs \( \{ \mathcal{M}(\bm{Z}^{(i)}) \}_{i=1}^{N_q} \) and computes statistical quantities (e.g., mean and variance).
\end{enumerate}
This design cleanly separates uncertainty propagation logic from the numerical solver implementation. 
As a result, all existing LBM applications in OpenLB (also beyond CFD) can be used without modification, and new UQ methods can be integrated into the framework by extending the modular interfaces of the UQ layer.

\section{Core class design} \label{sec:core_class_design}

The core of the OpenLB-UQ framework is organized as a collection of modular C++ classes, structured around abstract base classes and template-based specializations. 
This design enables flexibility in composing different UQ strategies and facilitates seamless integration of new methods.

Figure~\ref{fig:uml} presents the class hierarchy underlying the UQ layer. 
The framework is centered around the \texttt{UncertaintyQuantification} manager class, which coordinates all stages of the UQ process, including sampling, solver invocation, and postprocessing. 
Its functionality is composed of the following key components.

\subsection{Distribution and polynomial basis}

The \texttt{Distribution} class hierarchy defines input uncertainty models. 
It supports common univariate distributions such as uniform and Gaussian, as well as joint multivariate distributions constructed through product or correlated measures. 
Each distribution instance provides sampling functions and moments necessary for stochastic analysis.

To support spectral methods such as SC-gPC, each distribution is associated with an orthogonal polynomial basis through the \texttt{PolynomialBasis} interface. 
Concrete implementations like \texttt{LegendreBasis} and \texttt{HermiteBasis} correspond to uniform and Gaussian inputs, respectively. 
This automatic pairing guarantees consistency between sampling and projection in gPC expansions.

\subsection{Sampling strategies}

The \texttt{SamplingStrategy} interface defines a general contract for generating sample points in the stochastic space. 
It is implemented by various non-intrusive UQ methods, including:
\begin{itemize}\setlength{\itemsep}{0.0em}
    \item \texttt{MonteCarloSampling}
    \item \texttt{QuasiMonteCarloSampling}
    \item \texttt{LatinHypercubeSampling}
    \item \texttt{StochasticCollocation}
\end{itemize}
Each strategy class produces a sample matrix \( \{ \bm{Z}^{(i)} \}_{i=1}^{{N_q}} \subseteq \mathbb{R}^{d_Z} \), which is passed to the deterministic solver. 
For sampling-based methods, samples are drawn from the input distribution.
For collocation-based methods, points are selected via tensor-product or sparse-grid quadrature rules.

\subsection{Quadrature and collocation}

The \texttt{QuadratureRule} component handles the numerical integration over the stochastic space. 
This base class defines the interface for evaluating points and weights. 
Specific rules, including Gauss--Quadrature, Genz--Keister, and Clenshaw--Curtis, are implemented as subclasses~\cite{genz1996fully, trefethen2008gauss}.
The multi-dimensional quadrature is constructed via tensor products or sparse Smolyak grids~\cite{garcke2012sparse}.
The selection of quadrature rules is managed through a type-safe enumerator \texttt{QuadratureMethod}, which ensures consistency in point generation and projection.

For SC, the class \texttt{StochasticCollocation} evaluates the deterministic solver at quadrature points and computes the gPC coefficients via weighted projection or least-squares regression.

\subsection{Solver orchestration and data management}

The \texttt{UncertaintyQuantification} class acts as the top-level orchestrator. It holds instances of \texttt{SamplingStrategy}, \texttt{Distribution}, and \texttt{QuadratureRule}, and manages all simulation inputs and outputs. For each sample, it launches the solver, collects the quantity of interest, and updates statistical aggregates.

To interface with existing OpenLB applications, the UQ layer requires only a minimal wrapper that connects input samples to solver parameters and extracts output quantities. This lightweight coupling preserves the modularity of both the UQ and LBM components.

\section{Runtime workflow} \label{sec:runtime_workflow}

The execution of OpenLB-UQ follows a modular, three-phase runtime workflow: (i) sample generation, (ii) deterministic simulation, and (iii) statistical postprocessing. 
This structure enables a clear separation between the UQ logic and the numerical solver. 
Algorithm~\ref{alg:uq_openlb} provides a high-level overview of this process. 
\begin{algorithm}[ht!]
    \caption{Non-intrusive UQ algorithm in OpenLB.}
    \label{alg:uq_openlb}
    \begin{algorithmic}[1]
        \State \textbf{Input:} 
            random input \( \bm{Z} \sim h(\bm{Z}) \), LBM solver \( \mathcal{S} \), number of samples \( {N_q} \)
        \State \textbf{Output:} 
            mean \( \mathbb{E}[\mathcal{M}] \) and standard deviation \( \sigma(\mathcal{M}) \) of the quantity of interest \( \mathcal{M} \)
        \item[]
        \State Generate sample set \( \{\bm{Z}^{(i)}\}_{i=1}^{{N_q}} \) from \( h(\bm{Z}) \)
        \For{\( i = 1, \dots, N \)}
            \State \( \mathcal{M}^{(i)} \leftarrow \mathcal{S}(\bm{Z}^{(i)}) \) \Comment{Run LBM simulation with sample \( \bm{Z}^{(i)} \)}
        \EndFor

        \If{Monte Carlo sampling}
            \State \( \mathbb{E}[\mathcal{M}] \leftarrow \frac{1}{N_q} \sum_{i=1}^{N_q} \mathcal{M}^{(i)} \) \Comment{Eq.~\eqref{eq:meanMCS}}
            \State \( \sigma(\mathcal{M}) \leftarrow \sqrt{ \frac{1}{N_q} \sum_{i=1}^{N_q} \left( \mathcal{M}^{(i)} - \mathbb{E}[\mathcal{M}] \right)^2 } \) \Comment{Eq.~\eqref{eq:varMCS}}
        \EndIf
        
        \If{Stochastic collocation}
            \State Approximate: \( \mathcal{M}(\bm{Z}) \approx \sum_{k=0}^{N} \hat{\mathcal{M}}_k \, \Phi_k(\bm{Z}) \) \Comment{Eq.~\eqref{eq:gPC}}
            \State Compute coefficients: \( \hat{\mathcal{M}}_k \approx \frac{1}{\gamma_k} \sum_{i=1}^{N_q} a_i\, \mathcal{M}^{(i)}\, \Phi_k(\bm{Z}^{(i)}) \) \Comment{Eq.~\eqref{eq:coef_discrete}}
            \State \( \mathbb{E}[\mathcal{M}] \leftarrow \hat{\mathcal{M}}_0 \) \Comment{Eq.~\eqref{eq:gpc_mean}}
            \State \( \sigma(\mathcal{M}) \leftarrow \left( \sum_{k=1}^{N} \hat{\mathcal{M}}_k^2 \right)^{1/2} \) \Comment{Eq.~\eqref{eq:gpc_std}}
        \EndIf      
    \end{algorithmic}
\end{algorithm}
At runtime, the \texttt{UncertaintyQuantification} manager orchestrates the following steps:
\begin{enumerate}
    \item \textbf{Initialization:}  
    The UQ configuration is defined directly within the C++ application code. 
    This includes the choice of sampling strategy, distribution parameters, quadrature rule (if applicable), and the number of samples \( {N_q} \). 
    These parameters are passed to the \texttt{UncertaintyQuantification} manager.

    \item \textbf{Sample generation:}  
    A set of input samples \( \{ \bm{Z}^{(i)} \}_{i=1}^{N_q} \subset \mathbb{R}^{d_Z} \) is generated in the stochastic space using the selected \texttt{SamplingStrategy} instance. 
    Depending on the method, samples may come from random draws (e.g., Monte Carlo), low-discrepancy sequences (e.g., Sobol), or deterministic collocation points (e.g., Clenshaw--Curtis sparse grids).

    \item \textbf{Deterministic simulation:}  
    For each sample \( \bm{Z}^{(i)} \), a corresponding deterministic simulation input is assembled and passed to the OpenLB solver. 
    The solver executes a complete LBM run and computes a quantity of interest \( \mathcal{M}(\bm{Z}^{(i)}) \). 
    The UQ layer treats each run as a black box and stores only the scalar output \( \mathcal{M}^{(i)} \).

    \item \textbf{Postprocessing:}  
    The outputs \( \{ \mathcal{M}^{(i)} \}_{i=1}^{N_q} \) are aggregated after all simulations are completed. 
    For sampling-based methods, statistical moments such as the mean and standard deviation are computed directly from the ensemble. 
    For SC, the outputs are projected onto a polynomial chaos basis \( \{ \Phi_k \} \) using quadrature or regression to obtain the gPC coefficients \( \{ \hat{\mathcal{M}}_k \} \).

    \item \textbf{Output:}  
    Final results—including statistical summaries and surrogate models (e.g., gPC expansions)—are written to disk. 
    Optional postprocessing modules compute spatial statistics (e.g., pointwise mean and variance fields) and export them in \texttt{.vti} format for visualization in ParaView.
\end{enumerate}

This workflow is inherently parallelizable. 
Currently, two levels of parallelism are supported: (i) \emph{sample-level}, and (ii) \emph{domain-level}. 
Sample-level parallelism distributes simulation runs across available cores or nodes, while domain-level parallelism is managed internally by OpenLB using MPI. 
This hybrid parallelization strategy allows the UQ framework to scale seamlessly from local workstations to high-performance clusters with minimal configuration changes.

Performance results and scalability analysis are provided in Section~\ref{sec:perf}.

\section{Numerical experiments} \label{sec:results}

To demonstrate the effectiveness and applicability of the proposed OpenLB-UQ framework, we present results from benchmark simulations of a 2D flow past a circular cylinder as well as a 2D Taylor--Green vortex flow, both with parametric uncertainties up to dimension four. 
Relative error metrics are used to determine the convergence behavior with respect to references of high polynomial order.

The errors are then compared for various numbers of samples to approve individual accuracy orders of MCS-, QMC- and SC LBM, respectively. 
Finally, we validate the convergence of our framework in terms of a Wasserstein metric when computing statistical solutions with four-dimensional uncertainty.

\subsection{Flow past a circular cylinder with uncertain inlet velocity}\label{cylinder}
We consider a standard benchmark of incompressible flow past a circular cylinder, extended to account for uncertainty in the inlet velocity.  
The inlet velocity is modeled as
\[
    u_{\text{in}}(\xi) = 0.2 + 0.04\,\xi, \qquad \xi \sim \mathcal{U}(-1,1),
\]
which enters a parabolic inflow profile and induces variability in the Reynolds number and drag coefficient.

The domain is a two-dimensional channel of size \(22D \times 4D\), with a cylinder of diameter \(D\) placed near the inlet and vertically centered.  
The inlet velocity profile is prescribed as
\begin{align}
    \bm{u}(0,y,t,\xi) 
    = \begin{pmatrix}
        4\,u_{\text{in}}(\xi)\,\dfrac{y(H-y)}{H^2} \\
        0
      \end{pmatrix},
    \quad y \in [0,H],\; t \ge 0,\quad H=4D.
\end{align}

The Reynolds number is defined using the peak inlet velocity:
\[
    Re = \frac{u_{\text{in}} D}{\nu},
\]
with a nominal value of \( Re = 20 \) corresponding to \( u_{\text{in}} = 0.2 \).  

Boundary conditions are set as follows. At the inlet, an interpolated velocity boundary condition is applied~\cite{latt2008straight}; at the outlet, an interpolated pressure condition ensures free outflow~\cite{latt2008straight}.  
The cylinder surface is treated with a second-order accurate no-slip condition using the Bouzidi scheme~\cite{bouzidi2001momentum}, and the top and bottom walls are set to no-slip.  
The simulation is initialized with equilibrium populations, and the inflow is gradually ramped up to avoid transient oscillations.

The drag and lift forces are computed as surface integrals over the cylinder boundary \(S\),
\begin{align} 
    F_{\mathrm{D}} &= \int_S \left( -p\,n_1 + \rho \nu\, \frac{\partial v_t}{\partial \bm{n}}\, n_2 \right) \, \mathrm{d}S , \label{eq:dragForce} \\
    F_{\mathrm{L}} &= \int_S \left( -p\,n_2 - \rho \nu\, \frac{\partial v_t}{\partial \bm{n}}\, n_1 \right) \, \mathrm{d}S , \label{eq:liftForce}
\end{align}
where \(\bm{n} = (n_1, n_2)^{\mathrm{T}}\) is the outward unit normal vector on \(S\), and \(v_t = \bm{u} \cdot \bm{s}\) is the tangential velocity, with the unit tangent vector defined as \(\bm{s} = (s_1, s_2)^{\mathrm{T}} = (n_2, -n_1)^{\mathrm{T}}\). The derivative \(\partial v_t / \partial n\) denotes the directional derivative along \(\bm{n}\), i.e., \(\bm{n} \cdot \nabla v_t\).

The drag and lift coefficients are defined as
\begin{align}
    C_{\mathrm{D}} &= \frac{2 F_{\mathrm{D}}}{\rho\, u_{\text{in}}^{2}\, D} , \\
    C_{\mathrm{L}} &= \frac{2 F_{\mathrm{L}}}{\rho\, u_{\text{in}}^{2}\, D} ,
\end{align}

Let \(C_{\mathrm{D}}^{\mathrm{ref}}\) denote a reference value of the mean drag coefficient, obtained from a high-fidelity computation. Given an estimated mean \(\bar{C}_{\mathrm{D}}\), the relative error is computed as
\begin{equation}
    \delta := \frac{\left| C_{\mathrm{D}}^{\mathrm{ref}} - \bar{C}_{\mathrm{D}} \right|}
                   {\left| C_{\mathrm{D}}^{\mathrm{ref}} \right|} .
\end{equation}

This quantity is reported in decimal units throughout the results.

Before performing UQ, we need a deterministic solver that computes the drag coefficient for a given inlet velocity. In OpenLB, setting up a flow solver involves defining the computational domain, specifying boundary conditions, and executing the time-stepping loop until convergence. The function \texttt{simulateCylinder(...)} encapsulates this process.
A typical OpenLB setup follows these steps \cite{olbug:17}:
\begin{enumerate}
    \item Initialize the unit conversion: \\
        The \texttt{UnitConverter} handles transformations between physical and lattice units. 
        It is initialized based on the resolution and relaxation time.
        \begin{lstlisting}[language=C++]
UnitConverterFromResolutionAndRelaxationTime<T, DESCRIPTOR> const converter(...);
        \end{lstlisting}
    \item Prepare the computational geometry: \\
        The computational domain, including the flow boundaries and obstacle, is defined using the \texttt{SuperGeometry} class.
        \begin{lstlisting}[language=C++]
SuperGeometry<T,2> superGeometry(...);
        \end{lstlisting} 
    \item Prepare the LBM grid: \\
        The flow is simulated on a discrete lattice. 
        A \(D2Q9\) lattice is commonly used.
        \begin{lstlisting}[language=C++]
SuperLattice2D<T, DESCRIPTOR> lattice(...);
prepareLattice(...);
        \end{lstlisting}
    \item Time-stepping loop: \\
        The solver iterates over time steps until the solution reaches a steady state or periodic behavior. Boundary conditions are also enforced at each time step.
        \begin{lstlisting}[language=C++]
for (std::size_t iT = 0; iT <= converter.getLatticeTime(maxPhysT); ++iT) {
    // Update boundary conditions
    setBoundaryValues(...);
                
    // Execute collide and stream step
    lattice.collideAndStream();
                
    // Compute and output results at regular intervals
    if (iT % converter.getLatticeTime(.1) == 0) {
        drag = getResults(...);
    }
}
        \end{lstlisting}
\end{enumerate}

After setting up the deterministic solver, we integrate it into the UQ framework. 
A function \texttt{simulateCylinder(...)} internally executes the steps above while allowing variations in the inlet velocity for different UQ samples.

In the following, we document how the uncertainty in the inlet velocity is incorporated into the cylinder flow simulation. 
The quantity of interest is the drag coefficient on the cylinder surface, which is sensitive to variations in the inflow velocity. 
By performing multiple simulations (referred to as samples), we can capture how uncertain inlet conditions affect the overall flow behavior, particularly the drag coefficient.

To account for uncertainty in the inlet velocity, we specify a uniform distribution over a user-defined range. 
In code, we define this range with:
\begin{lstlisting}[language=C++]
auto dist = uniform(0.8 * u0, 1.2 * u0);
\end{lstlisting}
This object, \texttt{dist}, specifies that our inlet velocity is drawn from a uniform distribution with minimum \texttt{0.8 * u0} and maximum \texttt{1.2 * u0}.

Depending on the desired approach, the user can choose either MCS or a SC-gPC method for the UQ:
\begin{itemize}
    \item \textbf{MCS:} 
    This approach samples the distribution multiple times (denoted by \texttt{N}) to gather a statistical representation of the drag coefficient. 
    A seed value can be used for reproducibility:
    \begin{lstlisting}[language=C++]
UncertaintyQuantification uq(UQMethod::MonteCarlo);
unsigned int seed = 123456;
uq.initializeMonteCarlo(N, dist, seed);
    \end{lstlisting}

    \item \textbf{SC-gPC:} This method uses polynomial expansions to more efficiently approximate the dependence of the drag coefficient on the inlet velocity:
    \begin{lstlisting}[language=C++]
UncertaintyQuantification uq(UQMethod::GPC);
uq.initializeGPC(order, N, dist);
    \end{lstlisting}
    The parameter \texttt{order} refers to the polynomial order in the generalized polynomial chaos expansion, and the parameter \texttt{N} refers to the quadrature points used in each uncertainty dimension.
\end{itemize}

After setting up the desired UQ method, the user extracts the actual sample points (inlet velocity values) to be used in the flow solver:
\begin{lstlisting}[language=C++]
auto samples = uq.getSamplingPoints();
\end{lstlisting}
Each element \texttt{samples[n][0]} corresponds to one instance of the inlet velocity for the \texttt{n}th simulation.

\paragraph{Running the flow solver}
With the sample velocities at hand, we can run the cylinder flow solver for each sample. 
In this test case, the solver computes the flow around the cylinder on a mesh of resolution \texttt{resolution}, and outputs the drag coefficient:
\begin{lstlisting}[language=C++]
for (size_t n = 0; n < samples.size(); ++n) {
    // Run the cylinder simulation with the given inlet velocity
    dragCoefficients[n] = simulateCylinder(
        resolution,
        samples[n][0],
        gplot,
        writeVTM
    );
}
\end{lstlisting}

Once all simulations are completed, the drag coefficients which are stored in \texttt{dragCoefficients} can be processed to compute the mean and standard deviation, thereby quantifying the statistical response of the flow to uncertain inflow conditions:
\begin{lstlisting}[language=C++]
double meanDrag = uq.mean(dragCoefficients);
double stdDrag  = uq.std(dragCoefficients);
\end{lstlisting}

Initially, we perform a stochastic consistency study for SC-gPC on the 2D cylinder flow described above. 
This evaluation is based on the relative error, denoted as $\delta$, of the mean and standard deviation of drag coefficient $C_{\mathrm{D}}$. 
We investigate this across different spatial resolutions (here $n_x=10, 20, 40, 80$) and polynomial orders ($N=1, 2, 3, \ldots, 10$), quadrature points (${N_q} = 2N + 1 $). 
Hence, we compute 
\begin{align}
    \delta &= \frac{\left\vert \bar{C_{\mathrm{D}}}^{n_x} - \bar{C_{\mathrm{D}}}^{nx}_{10} \right\vert}{\left\vert \bar{C_{\mathrm{D}}}^{n_x}_{10}(t) \right\vert}, \label{eq:stochErrExp}\\
    \delta &= \frac{\left\vert \sigma\left({C_{\mathrm{D}}}^{n_x}_{N}(t)\right) - \sigma\left({C_{\mathrm{D}}}^{n_x}_{10}(t)\right) \right\vert}{\left\vert \sigma\left({C_{\mathrm{D}}}^{n_x}_{10}(t)\right) \right\vert},
\end{align}
respectively.

\begin{figure}[ht!]
    \centering
    \includegraphics[width=0.8\linewidth]{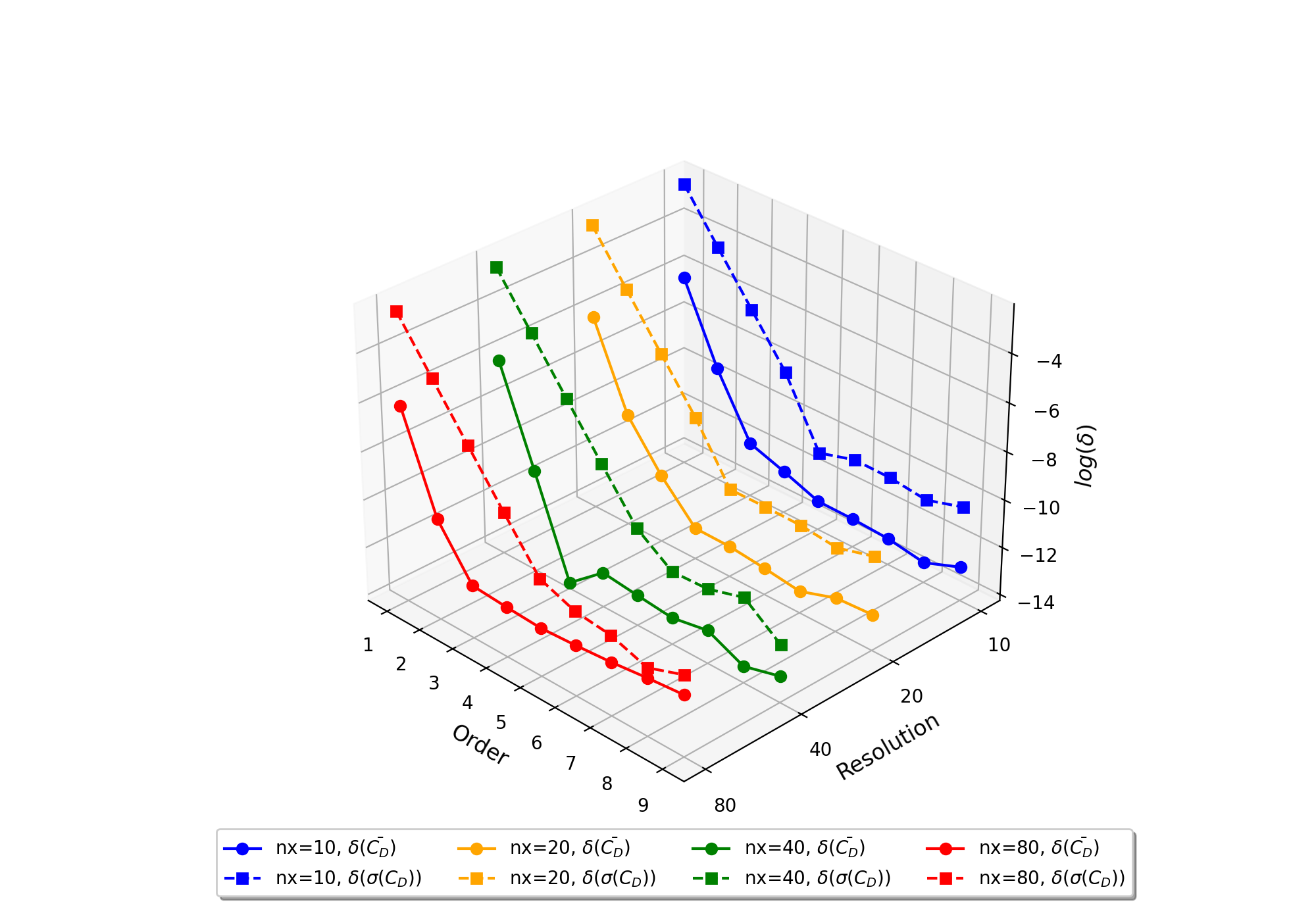}
    \caption{Relative error ($\delta$) of expectation value ($\bar{C_{\mathrm{D}}}$) and standard deviation ($\sigma(C_{\mathrm{D}})$) of drag coefficient \(C_{\mathrm{D}}\) for two-dimensional cylinder flow computed with SC-gPC with respect to highest polynomial order result. 
      Several spatial resolutions (\(n_x=10, 20, 40, 80\)) and polynomial orders (\(k=1,2,3,\ldots, 10\)) are tested.}
    \label{fig:cylinder2dConvergency}
\end{figure}

The convergence results in terms of the obtained relative error over several resolutions and polynomial orders are presented in Figure~\ref{fig:cylinder2dConvergency} for two dedicated points in time. The results show that a polynomial order of $k = 5$ is sufficient for achieving the highest accuracy.

\begin{figure}[ht!]
    \centering
    \subfloat[Mean velocity magnitude, $\bar{|\bm{u}|}$]{%
        \includegraphics[width=0.9\textwidth]{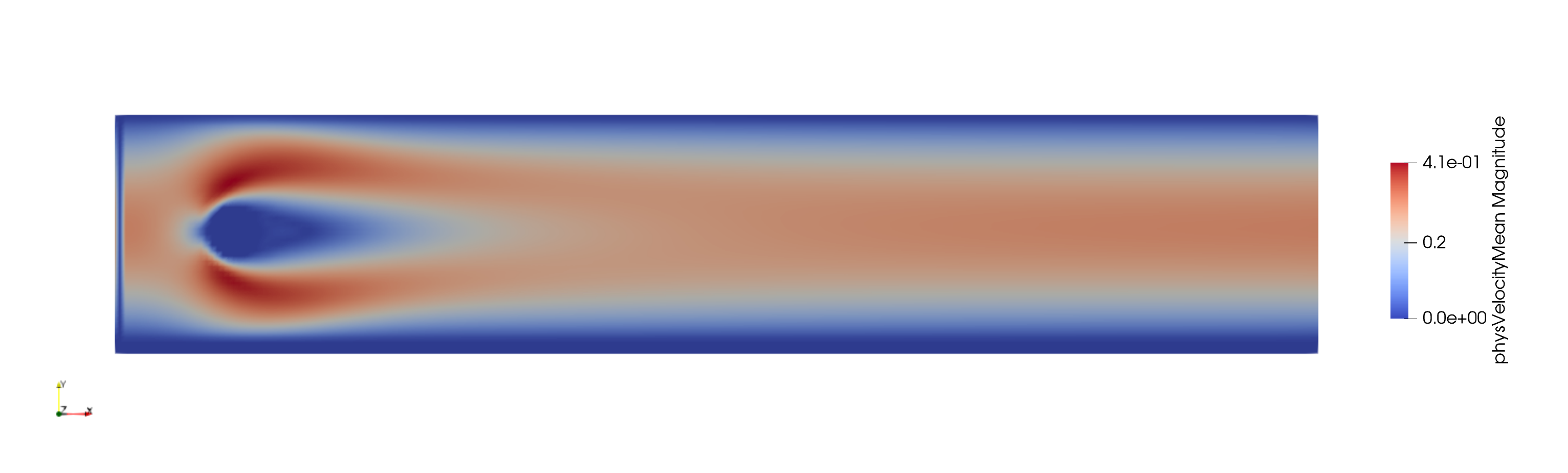} 
    }\\
    \subfloat[Standard deviation of velocity magnitude, $\sigma(|\bm{u}|)$]{%
        \includegraphics[width=0.9\textwidth]{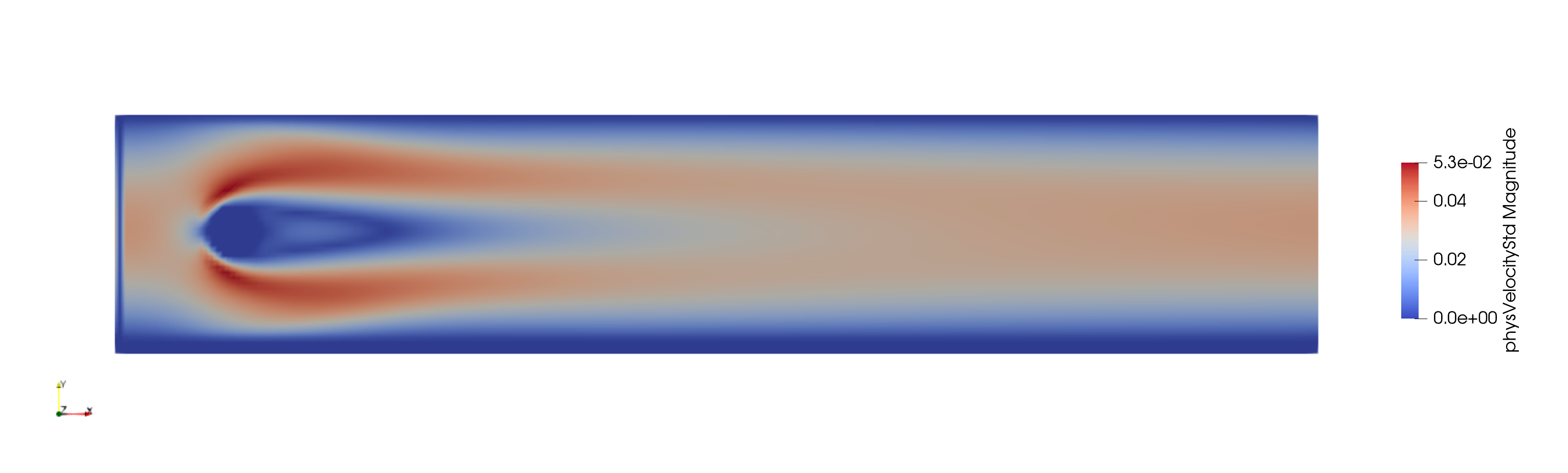}
    }
    \caption{Mean (\(\bar{|\bm{u}|}\)) and standard deviation (\(\sigma(|\bm{u}|)\)) of the velocity magnitude in the two-dimensional cylinder flow, computed using SC-gPC with a 5th-order polynomial expansion and 11 quadrature points (\(N = 11\)).}
    \label{fig:velocity_stats}
\end{figure}

From Figure~\ref{fig:cylinder2dConvergency}, it can be observed that the relative errors converge exponentially to machine precision. 
This indicates that here the SC-gPC achieves spectral accuracy in the random space.
The mean and standard deviation of the velocity magnitude of the flow field, computed using a 5th-order SC-gPC method with 11 quadrature points (\( N = 11 \)), are shown in Figure~\ref{fig:velocity_stats}. 

\begin{figure}[ht!]
  \centering
    \includegraphics[width=0.6\linewidth]{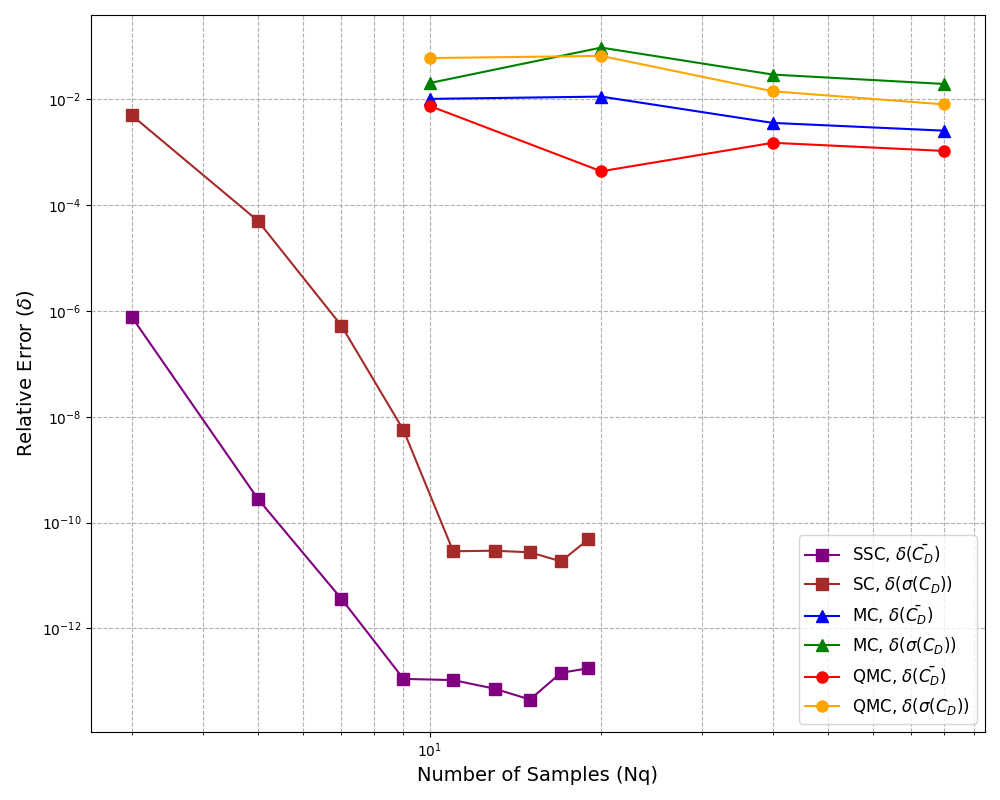}  
  \caption{Relative error of the expectation value (\(\delta(\bar{C_{\mathrm{D}}})\)) and standard deviation (\(\delta(\sigma(C_{\mathrm{D}}))\)) of the drag coefficient for the two-dimensional cylinder flow at \( n_x = 20 \). The Monte Carlo sampling (MCS) and quasi Monte Carlo (QMC) results are computed with different sample sizes (\( N = 10, 20, 40, 80, 100 \)) and compared against SC-gPC.}
  \label{fig:mc_vs_sc}
\end{figure}

Comparing MCS at resolution \( n_x = 20 \), we consider different numbers of samples (\( N = 10, 20, 40, 80, 100 \)). 
The results are compared with SC-gPC to analyze the convergence behavior. 
As shown in Figure~\ref{fig:mc_vs_sc}, the relative error of the estimated mean and standard deviation of the \( C_{\mathrm{D}} \) decreases as the number of MCS samples increases.

The SC-gPC results serve as a reference, demonstrating the efficiency of spectral approaches in capturing uncertainty with fewer samples.
\begin{figure}[ht!]
  \centering
    \includegraphics[width=0.6\linewidth]{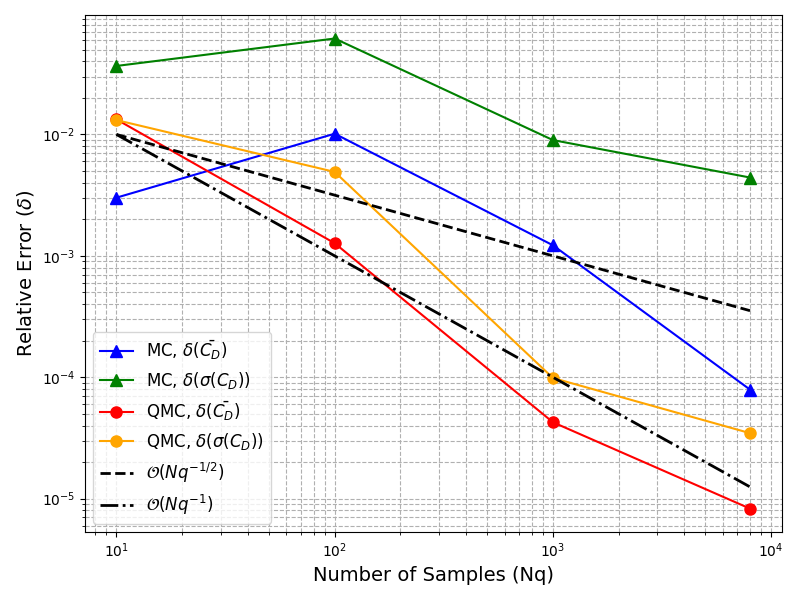} 
  \caption{Relative error of the expectation value \(\delta(\bar{C_{\mathrm{D}}})\) and standard deviation \(\delta(\sigma(C_{\mathrm{D}}))\) of the drag coefficient for the two-dimensional cylinder flow at \( n_x = 20 \). The Monte Carlo Sampling (MCS) and quasi Monte Carlo (QMC) results are computed with several sample sizes (\( N = 10, 100, 1000, 8000 \)).}
  \label{fig:mc_vs_qmc}
\end{figure}
\begin{table}[ht!]
  \centering
  \caption{Comparison of mean drag coefficient \(\bar{C_{\mathrm{D}}}\) and standard deviation \(\sigma(C_{\mathrm{D}})\) across different UQ methods with resolution $n_x = 20$.}
  \label{tab:uq_results}
  \begin{tabular}{lcccc}
    \hline
     & MC & QMC & SC-gPC & deterministic \\
    \hline
    \(\bar{C_{\mathrm{D}}}\) & 5.66731 & 5.67270 & 5.67261 & 5.63208 \\
    \(\sigma(C_{\mathrm{D}})\) & 0.34896 & 0.34890 & 0.34888 & -- \\
    \hline
  \end{tabular}
\end{table}

\subsection{Taylor--Green vortex flow with uncertain viscosity}

We proceed with the Taylor--Green vortex (TGV) flow in a two-dimensional periodic domain \(d_x = 2\), a classical benchmark problem for incompressible flow simulations. 
The TGV flow is fully periodic and admits a known analytical solution to the NSE, making it ideal for validating numerical methods by direct comparison with reference solutions. 

In this test case, we introduce uncertainty in the flow viscosity via a random Reynolds number. Specifically, we model the Reynolds number as a uniformly distributed random variable,
\[
\mathrm{Re} \sim \mathcal{U}(0.8\,\mathrm{Re}_0,\, 1.2\,\mathrm{Re}_0),
\quad\text{with }\mathrm{Re}_0 = 15.
\]
The corresponding kinematic viscosity is computed per realization as
\[
\nu = \frac{u_0 L}{\mathrm{Re}},
\]
which induces uncertainty in the decay rate of the flow. This allows us to quantify the resulting variation in the velocity field and its kinetic energy.
Statistical estimates (e.g., mean and standard deviation of velocity magnitude and kinetic energy) are computed using SC-gPC and MCS.

The exact velocity field \(\bm{u}(x,y,t)\) and pressure field \(p(x,y,t)\) are given by
\begin{align}
 \bm{u}(x, y, t)
 & = 
 \begin{pmatrix} 
    -u_{0} \cos\bigl(k_{x} x\bigr) \sin\bigl(k_{y} y\bigr) \mathrm{e}^{\frac{-t}{t_{d}}} \\
    u_{0} \sin\bigl(k_{x} x\bigr) \cos\bigl(k_{y} y\bigr) \mathrm{e}^{\frac{-t}{t_{d}}}
 \end{pmatrix}, 
 \label{eq:TGV_vel}\\[6pt]
 p(x, y, t)
 & =
 -\frac{1}{4} u_{0}^{2} \left[
   \cos\left(2 k_{x} x\right) + \left( \frac{k_{x}}{k_{y}}\right)^{2} \cos\left(2 k_{y}y\right)
  \right]
  \mathrm{e}^{\frac{-2t}{t_{d}}} + P_{0}.
 \label{eq:TGV_press}
\end{align}

In \eqref{eq:TGV_vel} and \eqref{eq:TGV_press}, \(u_0\) is the initial velocity amplitude, \(k_x\) and \(k_y\) are wavenumbers corresponding to the domain length in the \(x\) and \(y\) directions, and \(t_d\) is a characteristic decay time that depends on the fluid viscosity.

We define a square domain \(\Omega = [\,0, n_x\,]\times[\,0, n_y\,]\) in lattice units, where \(n_x\) and \(n_y\) denote the number of grid nodes in each direction. 
For simplicity, we let \(n_x = n_y\), and we set the physical domain size to \(L = 2\pi\). 
Hence, the wavenumbers become 
\begin{align}
  k_{x} = \frac{2\pi}{n_x}, \qquad
  k_{y} = \frac{2\pi}{n_y}.
\end{align}
The initial density is determined via the ideal equation of state \(\rho = p / c_s^2\), where \(c_s\) is the speed of sound. 
The characteristic decay time is defined by
\begin{align}
  t_d = \frac{1}{\nu,(k_x^2 + k_y^2)}
\end{align}
and the corresponding lattice Boltzmann relaxation time is given by
\begin{align}
  \tau = \frac{\nu}{c_s^2} + \frac{1}{2}.
\end{align}

All simulations for this test case enforce periodic boundary conditions in both spatial directions to obtain a periodic numerical solution. 
The first global statistical quantity that we study is the (normalized) total kinetic energy, defined as
\begin{align}
  K(t) 
  = 
  \frac{2}{\vert\Omega\vert\,u_0^2}
  \int_{\Omega} 
    \Bigl(u^2(x,y,t) + v^2(x,y,t)\Bigr)\,
  \mathrm{d}x\,\mathrm{d}y,
  \label{eq:KineticEnergy}
\end{align}
where \(\Omega\) is the two-dimensional computational domain \([0,n_x]\times[0,n_y]\).  The factor of \(2 / (\vert\Omega\vert u_0^2)\) ensures normalization by the domain area \(\vert \Omega\vert\) and the square of the initial velocity amplitude \(u_0\). 
In practice, we approximate the integral via a discrete sum over the lattice nodes, i.e.
\begin{align}
  K(t)
  \approx
  \frac{2}{n_x n_y u_0^2}
  \sum_{x=0}^{n_x-1}\sum_{y=0}^{n_y-1} 
    \bigl[ u^2(x,y,t) + v^2(x,y,t)\bigr].
\end{align}
The implementation of the case in OpenLB-UQ follows the same guidelines as explained in detail for the flow around a cylinder case in Section~\ref{cylinder}.

To assess convergence toward a reference solution \(K_{\mathrm{ref}}(t)\), we define the relative error in the mean kinetic energy as
\begin{align}
  \delta(t) 
  = \frac{\left\vert \overline{K}(t) - K_{\mathrm{ref}}(t) \right\vert}
         {\left\vert K_{\mathrm{ref}}(t) \right\vert},
  \label{eq:RelativeErrorK}
\end{align}
where \(\overline{K}(t)\) is the computed mean and \(K_{\mathrm{ref}}(t)\) is obtained from a finer spatial or stochastic discretization.  

In SC-gPC, accuracy depends on both the polynomial order \(k\) and the quadrature points number \(N\), since a deterministic solve is performed at each quadrature node.  
To study the effect of \(k\) alone, we fix \(N=1000\) and vary \(k\) from \(1\) to \(8\).  

\begin{figure}[ht!]
  \centering
  \subfloat[Expectation value errors $\delta(\bar{K}(0.5t_d))$]{\includegraphics[width=0.48\textwidth]{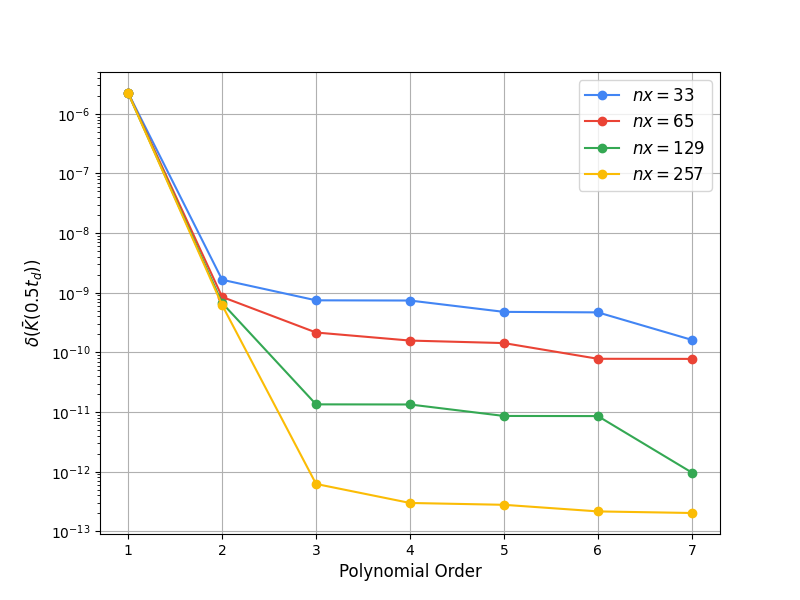}}
  \subfloat[Standard deviation errors $\delta(\sigma(K(0.5t_d)))$]{\includegraphics[width=0.48\textwidth]{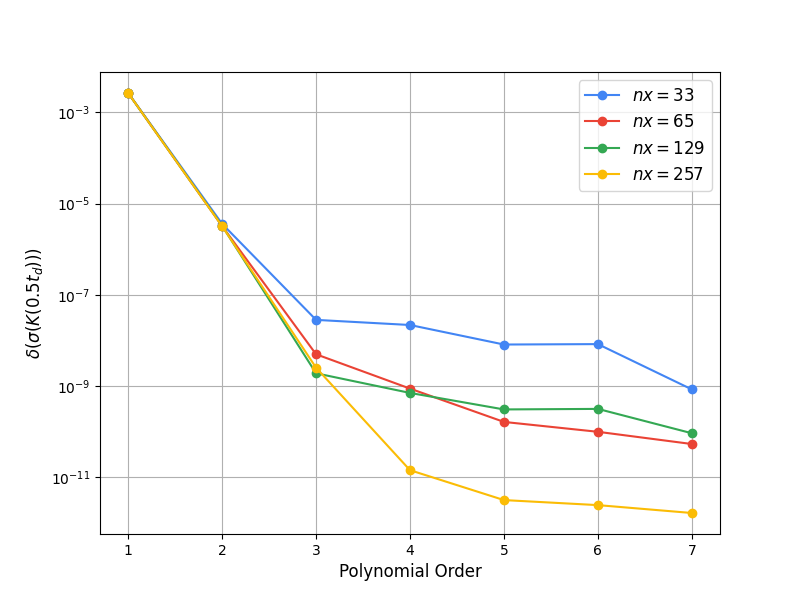}}  
  \caption{Relative error ($\delta(t)$, Eq~\eqref{eq:RelativeErrorK}) of expectation value $\bar{K}(t)$ and standard deviation $\sigma(K(t))$ of normalized total kinetic energy \(K(t)\) for TGV flow with uncertain viscosity computed with SC LBM with respect to highest polynomial order (\(k=8\)) results, respectively for individual resolutions at time \(t = 0.5 t_d\). 
  Several spatial resolutions (\(n_x=33, 65, 129, 257\)) and polynomial orders (\(N=1,2,3,\ldots, 8\)), quadrature points (\(N = 1000\)) are tested.}
  \label{fig:scConvergence}
\end{figure}

Figure~\ref{fig:scConvergence} shows \(\delta(t)\) for the mean \(\overline{K}(t)\) and the standard deviation \(\sigma(K(t))\) of the normalized total kinetic energy at \(t=0.5\,t_d\), measured against the highest polynomial-order solution.  
The error decreases exponentially with \(k\), confirming the spectral convergence of SC-gPC when \(N\) is sufficiently large.

Beyond a certain order (e.g., $k \approx 3$--$4$), the convergence curves flatten, indicating that the total error is dominated by the deterministic discretization in physical space.

In other words, at sufficiently high polynomial order, the SC approach becomes more accurate than the underlying spatial grid can represent, and hence the relative error plateaus at the level prescribed by the spatial discretization error.

Following the convergence analysis in Figure~\ref{fig:scConvergence}, we fix the polynomial order at \(k = 5\) and investigate the effect of the quadrature resolution \(N\) in SC-gPC, comparing it against MCS.

\begin{figure}[ht!]
  \centering
  \subfloat[Expectation value errors $\delta(\bar{K}(0.5t_d))$]{\includegraphics[width=0.48\textwidth]{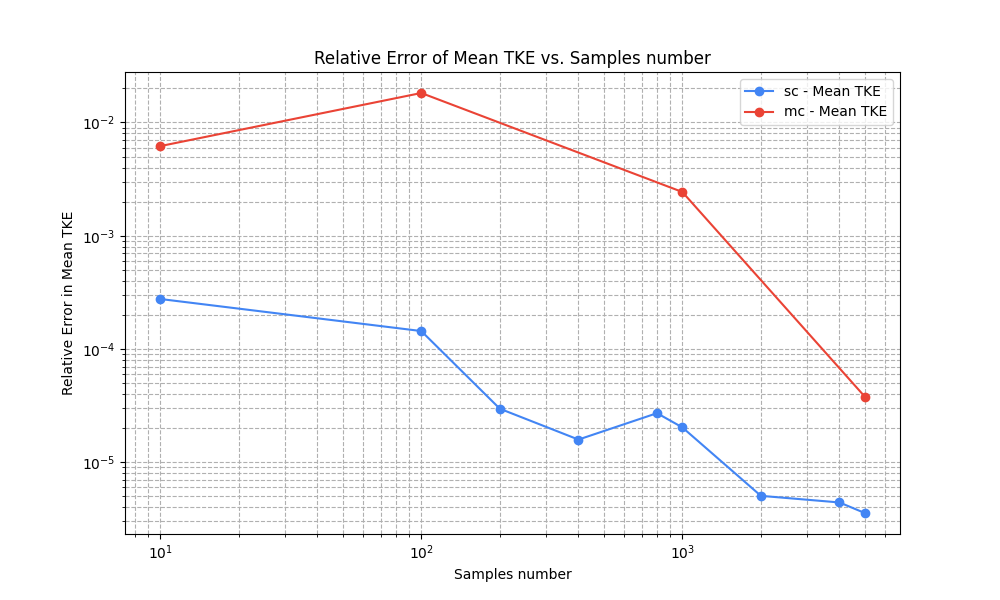}}
  \subfloat[Standard deviation errors $\delta(\sigma(K(0.5t_d)))$]{\includegraphics[width=0.48\textwidth]{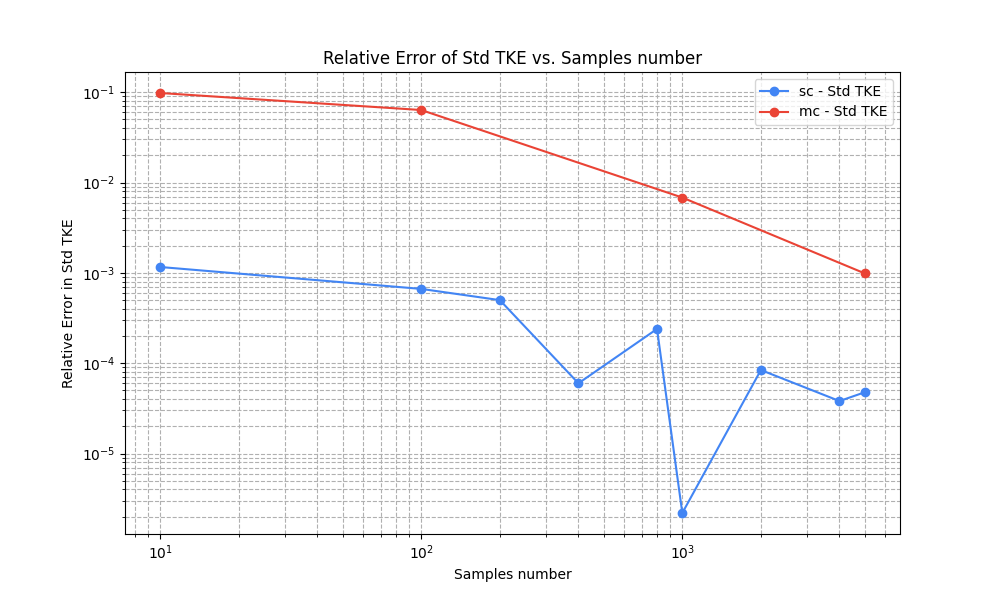}}
  
  \caption{Relative error ($\delta(t)$, Eq~\eqref{eq:RelativeErrorK}) of expectation value $\bar{K}(t)$ and standard deviation $\sigma(K(t))$ of normalized total kinetic energy \(K(t)\) for TGV flow with uncertain viscosity computed with SC LBM with respect to highest quadrature points number (\(N=10000\)) results at time \(t = 0.5 t_d\).
  Several spatial resolutions (\(n_x=33\)) and quadrature points (\(N=10,100,200,\ldots, 10000\)), Monte Carlo samples (\(N = 10, 100, 10000\)) are tested.}
  \label{fig:scConvergenceN}
\end{figure}

Figure~\ref{fig:scConvergenceN} compares SC-gPC (\(k=5\)) and MCS for the two-dimensional TGV flow at \(n_x = 33\), showing the relative error of the mean and standard deviation of \(K(t)\) at \(t=0.5\,t_d\) as functions of the number of quadrature points (SC-gPC) or samples (MCS).

For MCS, the errors decay at the expected \(O(N^{-0.5})\) rate for both mean and standard deviation.  
With SC-gPC at fixed polynomial order (\(k=5\)), increasing the number of quadrature points still yields a much faster convergence than MCS.

\subsection{Taylor--Green vortex flow with four-dimensional uncertain initial velocity perturbation}

This second version of the TGV flow is taken from~\cite{zhong2024stochastic, simonis2024computing}.
The velocity field and the pressure of the TGV are defined as 
\begin{align}\label{eq:taylorGreen}
 \bm{u} \left( x,y,t\right) 
 & = 
 \begin{pmatrix} u \left( x,y,t \right) \\ v \left( x,y,t \right) \end{pmatrix} 
  = 
 \begin{pmatrix} - {u_{0}^R} \cos\left( 2 x \right) \sin\left( 2 y \right) e^{-\frac{t}{t_{d}}} \\  {u_{0}^R} \sin\left( 2 x \right) \cos\left( 2 y \right)e^{\frac{t}{t_{d}}} \end{pmatrix}, \\
p\left( {x,y,t} \right) 
&= 
- \frac{1}{4}{u_{0}^R}^{2}\left[ \cos\left( 4 x \right) + \left( \frac{2}{2} \right)^{2} \cos\left( 4 y \right) \right] e^{-\frac{2t}{t_{d}}} + P_{0},
\end{align}
respectively, where, $x,y \in [0, 2\pi]$. 
The uncertain velocity ${u_{0}^R} = u_{0} + \epsilon_d(x)$ is based on the perturbation $\epsilon_d$ given by the first-order harmonics with four-dimensional uniformly i.i.d.\ random amplitude $\zeta_{d,i,j} \sim \mathcal{U}(-0.025, 0.025)$, i.e.
\begin{align}
\epsilon_d (x, y) = \frac{1}{4} \sum_{(i, j) \in \{0,1\}^2} \zeta_{d, i, j} k_i (4x) k_j (4y),
\end{align}
where 
\begin{align}
k_i (x) =
\begin{cases} 
\sin (x), & \text{if } i = 0, \\
\cos (x), & \text{if } i = 1.
\end{cases}
\end{align}
The four-dimensional uncertainty is straightforward to set up in our framework by defining a joint distribution as follows:
   \begin{lstlisting}[language=C++]
auto perturb1 = uniform(-0.025, 0.025);
auto perturb2 = uniform(-0.025, 0.025);
auto perturb3 = uniform(-0.025, 0.025);
auto perturb4 = uniform(-0.025, 0.025);
auto jointPerturb = joint({perturb1, perturb2, perturb3, perturb4});
   \end{lstlisting}

The maximum simulation time is $T_{\mathrm{max}} = 100$. 
Other parameter settings for the test case are described below and summarized in Table~\ref{tab:simulation_parameters}. 
The number of samples \(N\), the spatial resolution $n_{x}=n_{y}$, and the Reynolds number $Re$ scale linearly, while the Mach number $Ma$ scales inversely proportional. 
Based on the latter three parameters, the relaxation time $\tau$, the time step $\triangle t$, and the grid spacing $\triangle x$ are determined.
\begin{table}[ht!]
    \centering
    \caption{Simulation parameters for the TGV with four-dimensional uncertain initial condition.}
    \label{tab:simulation_parameters}
    \begin{tabular}{rrlrlll}
        \hline
          \makecell[c]{$N$} & \makecell[c]{$n_{x}=n_{y}$} &    \makecell[c]{$Ma$}    &   \makecell[c]{$Re$}    &   \makecell[c]{$\tau$}  &   \makecell[c]{$\triangle x$}  &   \makecell[c]{$\triangle t$}  \\ 
        \hline
        $8$   &   $8$   &   $0.2$     &   $320$     &   $0.501206$    &   $0.897598$    &   $0.103646$    \\
        $16$  &   $16$  &   $0.1$     &   $640$     &   $0.500646$    &   $0.418879$    &   $0.024184$    \\
        $32$  &   $32$  &   $0.05$    &   $1280$    &   $0.500334$    &   $0.202683$    &   $0.005850$  \\
        $64$  &   $64$  &   $0.025$   &   $2560$    &   $0.500170$    &   $0.099733$   &   $0.001439$  \\
        $128$ &   $128$ &   $0.0125$  &   $5120$    &   $0.500085$    &   $0.049473$   &   $0.000357$ \\
        $256$ &   $256$ &   $0.00625$ &   $10240$   &   $0.500043$    &   $0.024639$   &  $0.000088$ \\
        \hline
    \end{tabular}
\end{table}

\begin{figure}[ht!]
  \centering
  \subfloat[Deterministic]{\includegraphics[width=0.48\textwidth]{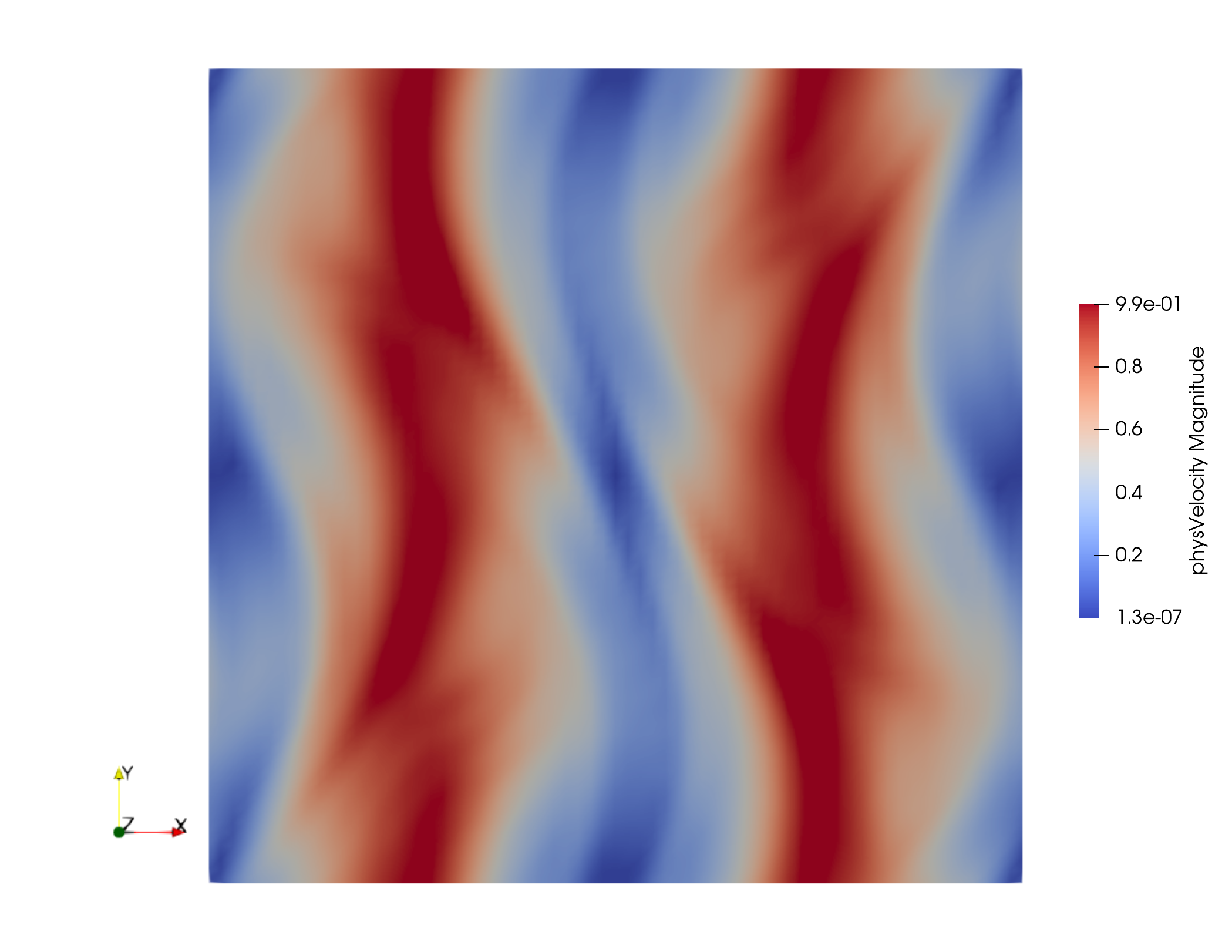}}
  \subfloat[Expected value]{\includegraphics[width=0.48\textwidth]{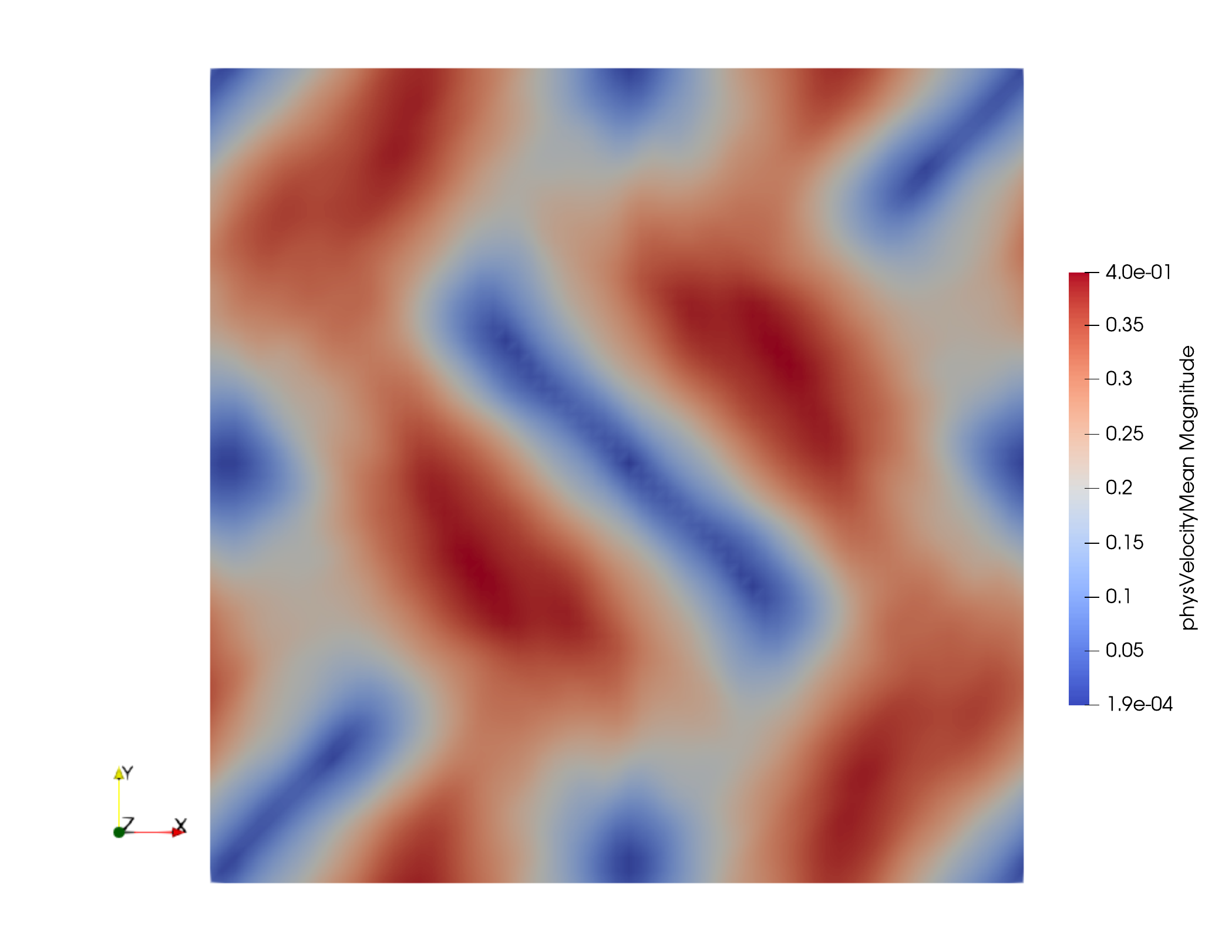}}
  
  \caption{Deterministic and expected velocity magnitude of the TGV with four-dimensional uncertain initial velocity computed with OpenLB-UQ for a resolution of $n = 256$ at $t = T_{\mathrm{max}}$.}
  \label{fig:detAndMeanTGV4d}
\end{figure}

Snapshots of the velocity magnitude computed from the TGV for deterministic and perturbed initial conditions (expected value in the latter case), respectively, are shown for \(t=T_{\mathrm{max}}\) in Figure~\ref{fig:detAndMeanTGV4d}. 

\begin{figure}[ht!]
\centering
\includegraphics[width=0.8\linewidth,trim={0 0 0 3cm},clip]{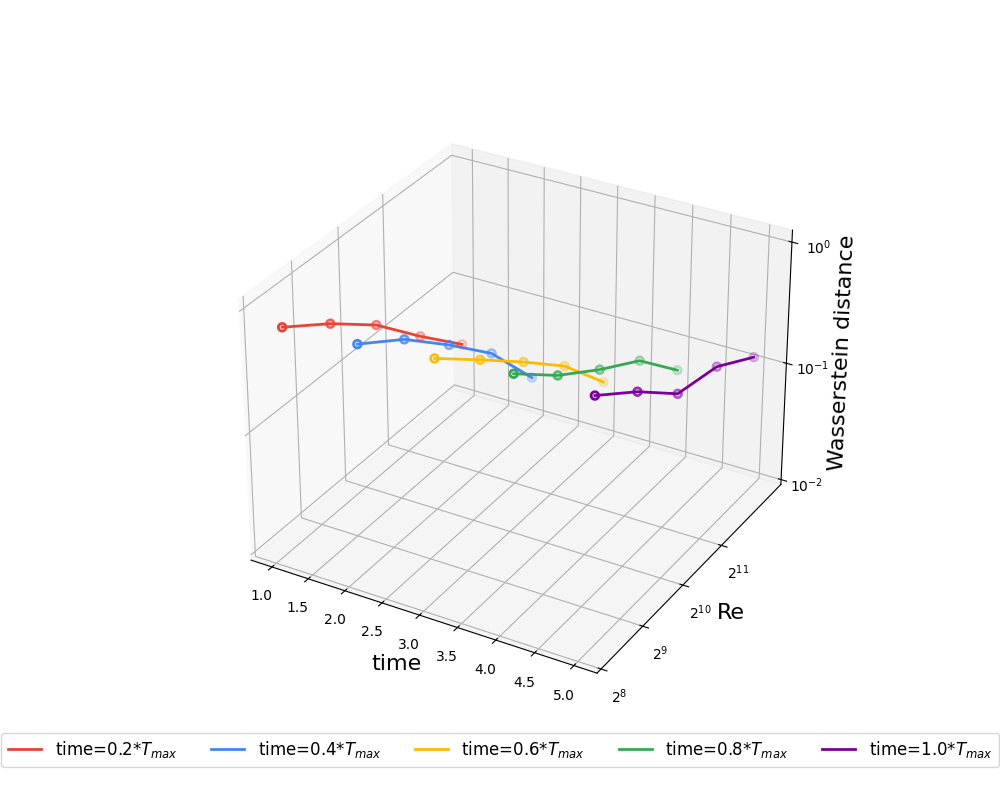}
\caption{\(1\)-Wasserstein distance at several time steps of the numerical statistical solutions with respect to the most resolved one (\(n_x=256\)) for the TGV with four-dimensional uncertain initial velocity computed with OpenLB-UQ.}
\label{fig:convergency3D}
\end{figure}
To evaluate the inviscid statistical convergence \cite{lanthaler20201statistical} of OpenLB-UQ, we approximate the $1$-Wasserstein distance defined as
\begin{equation}
    W_1(Q_1, Q_2) = \left( \inf_{\gamma \in \Gamma(Q_1, Q_2)} \int_{X^2} |x-y| \, d\gamma(x, y) \right),
\end{equation}
where $\Gamma(Q_1, Q_2)$ denotes the set of all joint distributions $\gamma$ on $M \times M$ whose marginals are $Q_1$ and $Q_2$ respectively, $d(x, y)$ is the distance between points $x$ and $y$ in the metric space $M$.
Here, \(Q_1\) and \(Q_2\) are numerical statistical solutions for resolutions \(n_x<256\) and the reference numerical statistical solution with resolution \(n_x=256\) (with matching parameters according to Table~\ref{tab:simulation_parameters}), respectively.

The time-dependent numerical statistical solutions are constructed by expectation values of Dirac measures from computed sample velocity fields at time \(t\). 
Hence, with each spatial resolution \(n_{x}\), we approximate the continuous statistical solutions to an initially perturbed NSE on \(\mathcal{X} = \mathbb{R}^{d_{x}}\) defined by probability measures on \(L^{2}_{\mathrm{div}}(\mathcal{X}; \mathcal{U})\) \cite{simonis2024computing}.

\begin{figure}[ht!]
  \centering
  \subfloat[$t = 0.2*T_{max}$]{\includegraphics[width=0.48\textwidth]{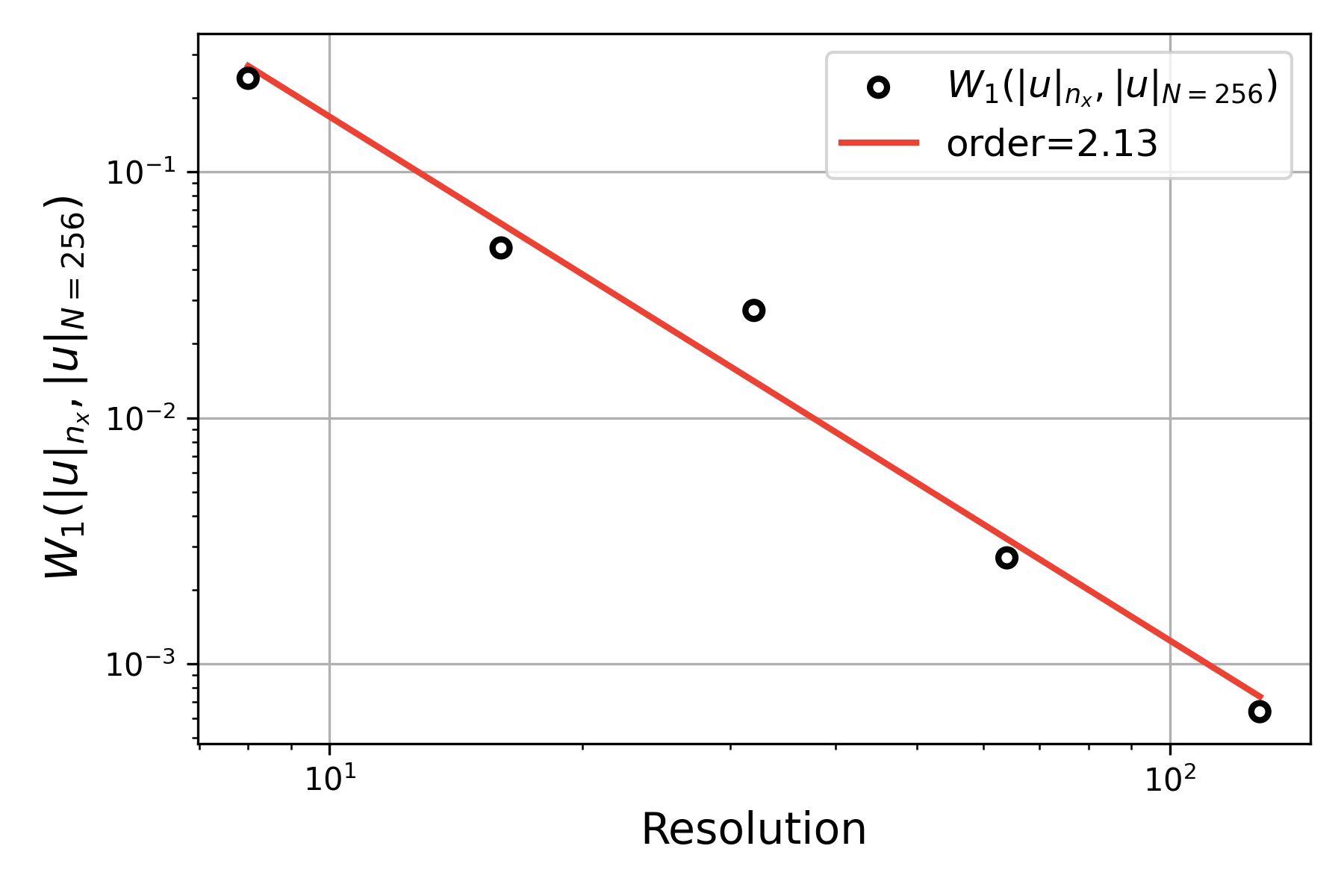}}
  \subfloat[$t = 0.4*T_{max}$]{\includegraphics[width=0.48\textwidth]{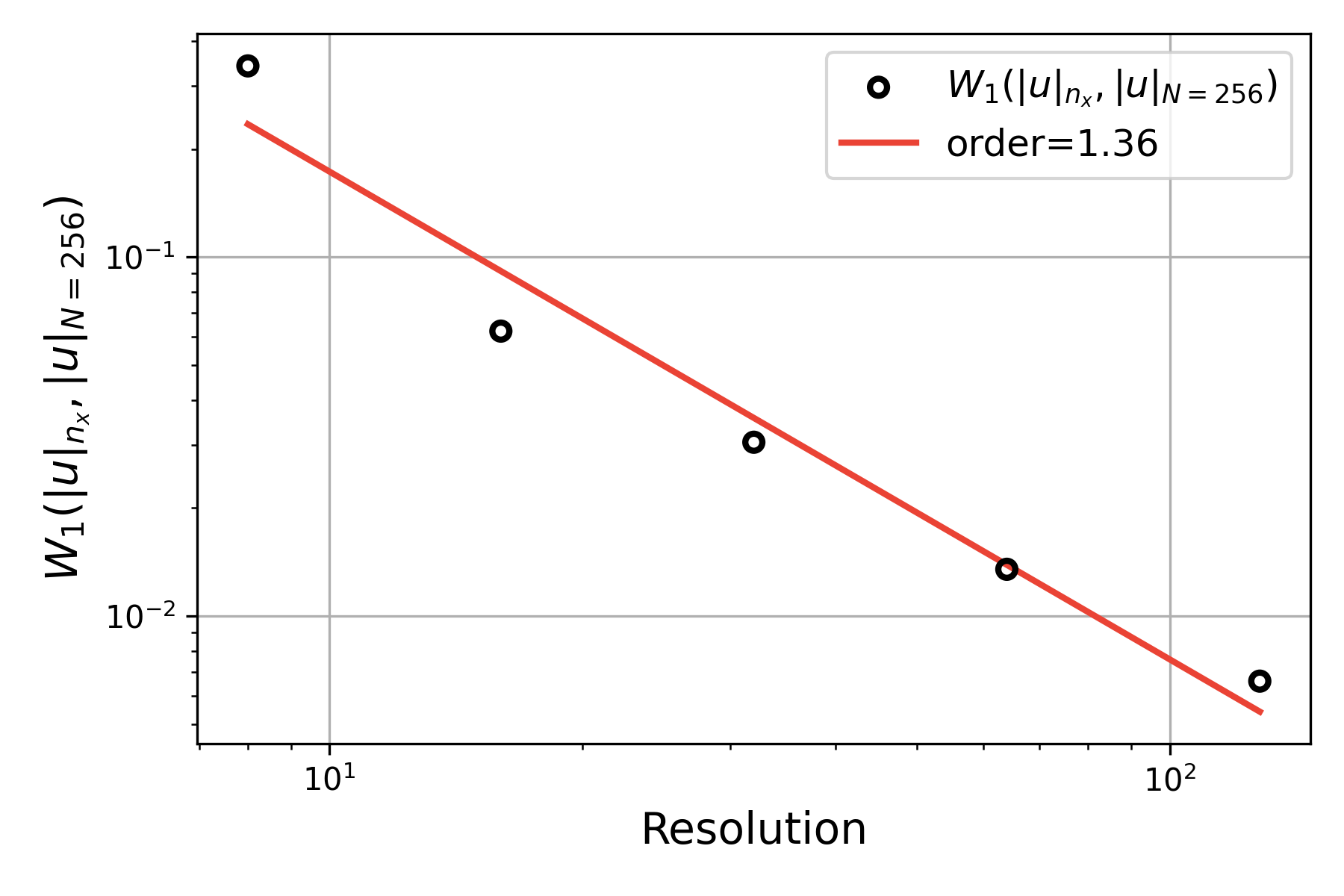}}\\
  \subfloat[$t = 0.6*T_{max}$]{\includegraphics[width=0.48\textwidth]{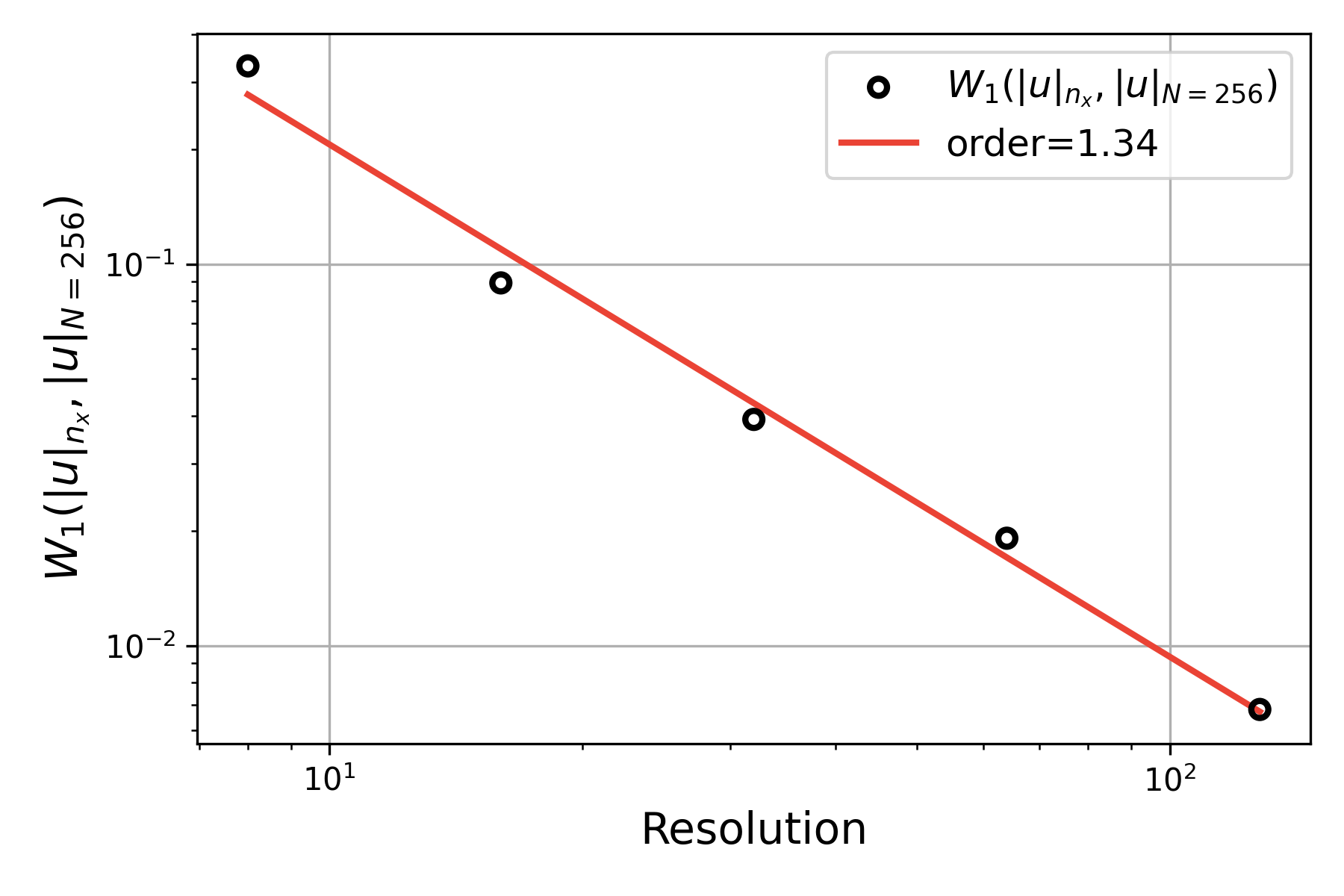}}
  \subfloat[$t = 0.8*T_{max}$]{\includegraphics[width=0.48\textwidth]{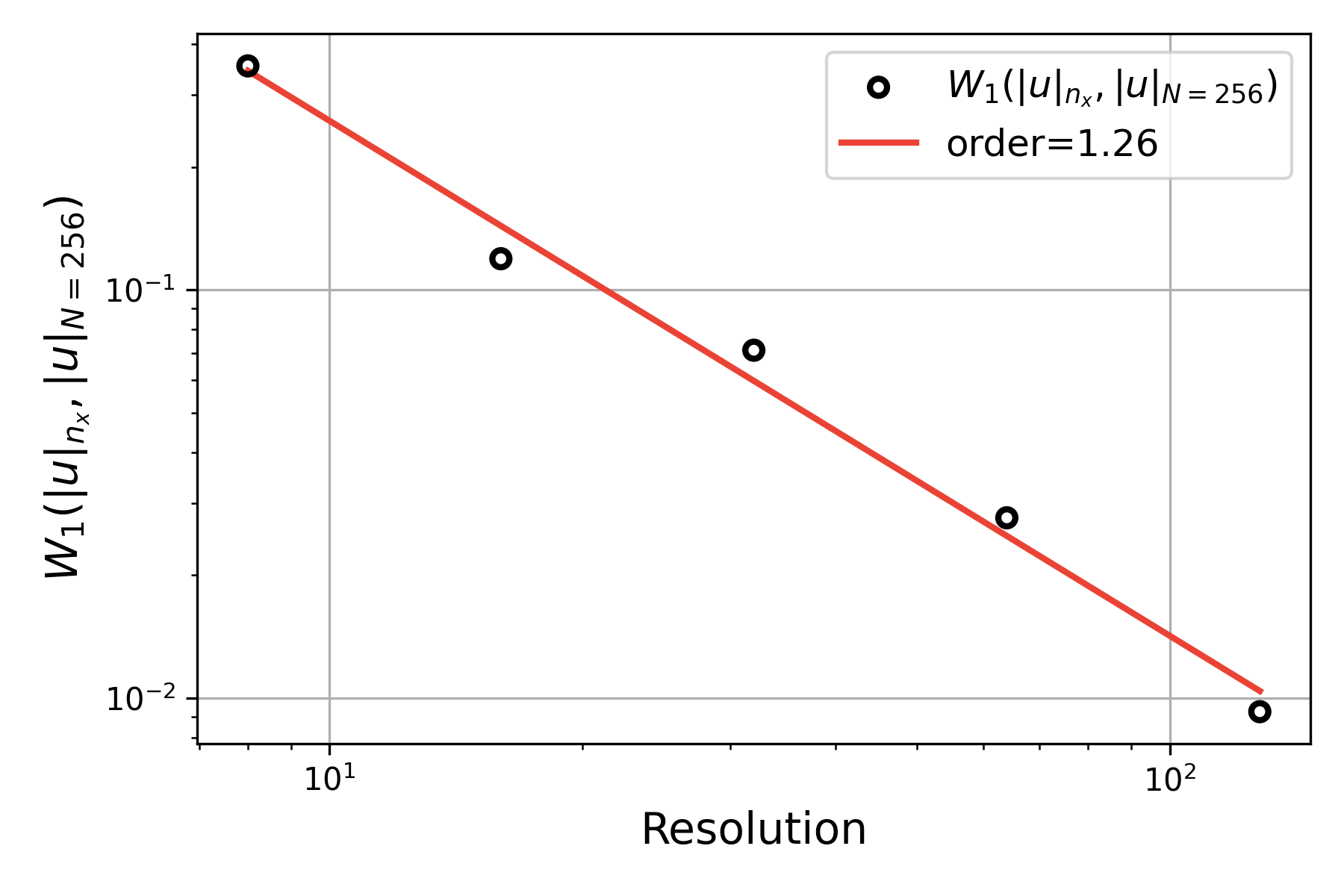}}\\
  \subfloat[$t = T_{max}$]{\includegraphics[width=0.48\textwidth]{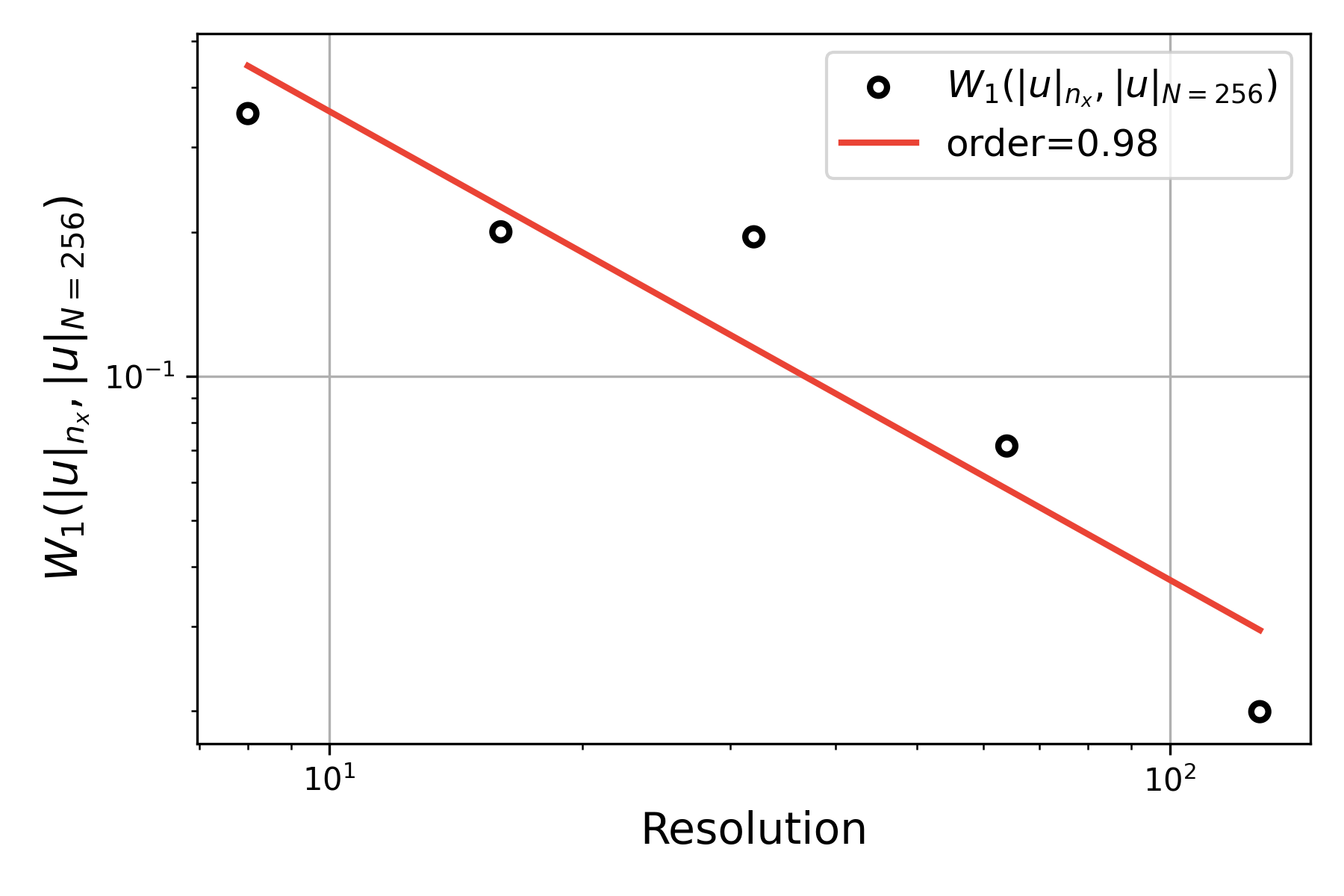}} 
  \subfloat[Convergence order over time]{\includegraphics[width=0.48\linewidth]{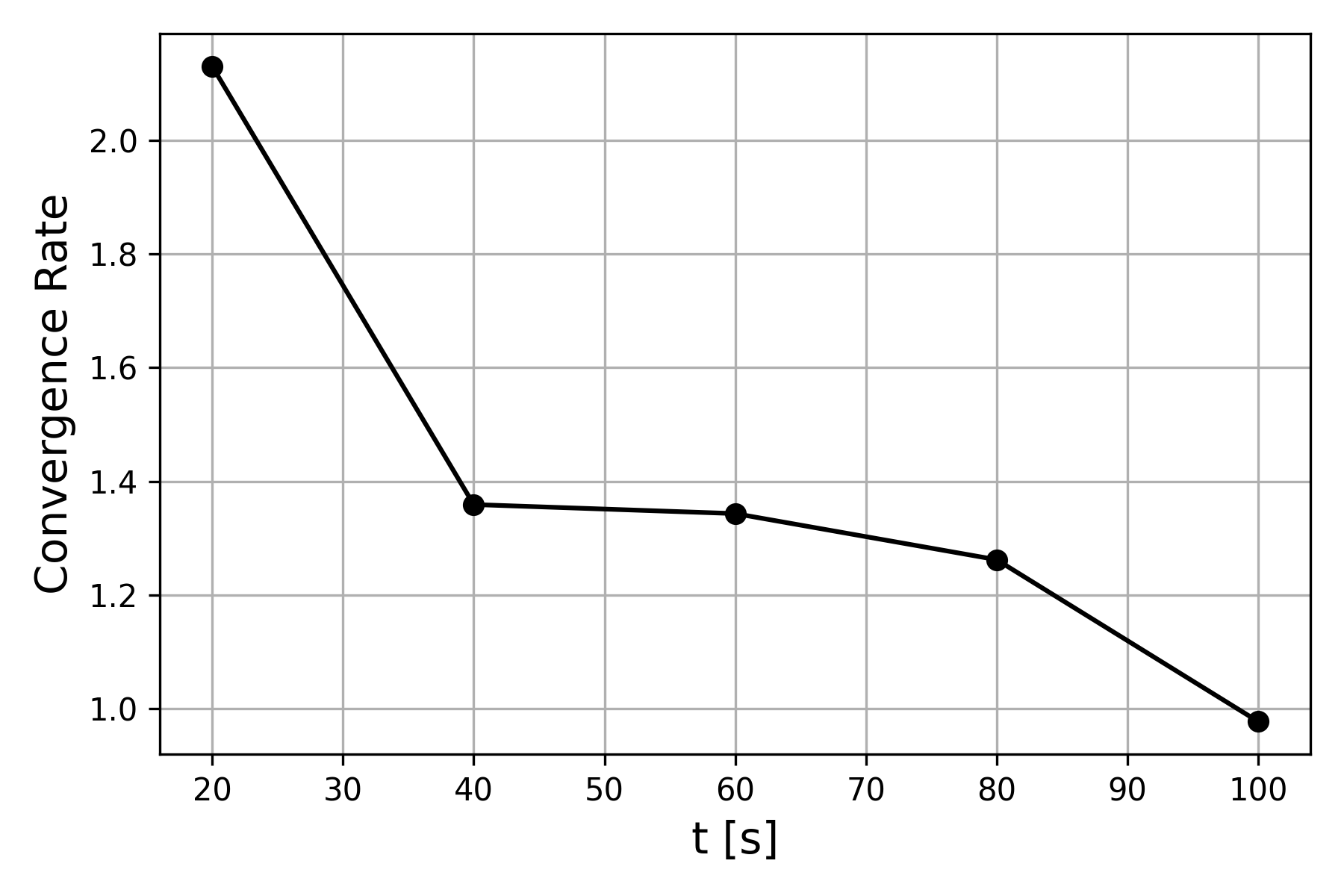} \label{fig:convergency_rate}}
  
  \caption{Convergence orders of \(1\)-Wasserstein distance at several time steps (a,b,c,d,e) and over time (f) of the numerical statistical solutions with respect to the most resolved one (\(n_x=256\)) for the TGV with four-dimensional uncertain initial velocity computed with OpenLB-UQ.
  }
  \label{fig:convergency}
\end{figure}

We approximate the Wasserstein distance from the computed sample data using the \texttt{scipy.stats.wasserstein\_distance} function in \texttt{SciPy}~\cite{2020SciPy-NMeth}.
If the approximated Wasserstein distance of the numerical statistical solutions for \( n = 8, 16, 32, 64, 128 \) to the most resolved one for \(n_{x}=256\), i.e.\ \( W_1(|u|_{n_{x}} (t), |u|_{N=256} (t)) \) converges, we thus indicate Wasserstein convergence towards a unique statistical solution of the incompressible Euler equations \cite{lanthaler20201statistical}.
The results indeed show that \( W_1(|u|_{n}, |u|_{N=256})\) converges as the Reynolds number increases with \(n_{x}\), cf.\ Figure~\ref{fig:convergency3D}.
The estimated convergence rates are provided in Figure~\ref{fig:convergency}. 
Note that the decrease of the Wasserstein convergence order over time toward a value of \(0.5\) (see Figure~\ref{fig:convergency_rate}) is expected for MCS.

\section{Performance evaluation}\label{sec:perf}

To evaluate the scalability of OpenLB-UQ, we conducted additional performance tests on the HoreKa supercomputer. 
Specifically, we utilized a CPU-only compute node with two Intel Xeon Platinum 8368 processors, offering a total of 76 physical cores (152 hardware threads) per node. Each processor operates at a base frequency of 2.40~GHz. 
All simulations were executed on a single node to focus exclusively on intra-node parallel efficiency.

To assess the practical benefits of different parallelization strategies, we compared the total simulation time required to process all 137 samples using two parallelization strategies:
\begin{enumerate}
    \item \textbf{Sample-level decomposition}: Each sample is executed independently on different processes or cores, without MPI-based domain decomposition within a single sample.
    \item \textbf{Domain-level decomposition}: Each sample runs in parallel using MPI across multiple cores with domain decomposition.
\end{enumerate}

First, we evaluate the scalability of the domain-level (MPI-based) parallelization strategy using the \texttt{cylinder2d} test case. 
For each resolution ($n_x=10$, $20$, $40$, and $80$), we perform strong-scaling tests by varying the number of MPI processes from 2 to 64. 
\begin{figure}[ht!]
  \centering
  \includegraphics[width=0.7\textwidth]{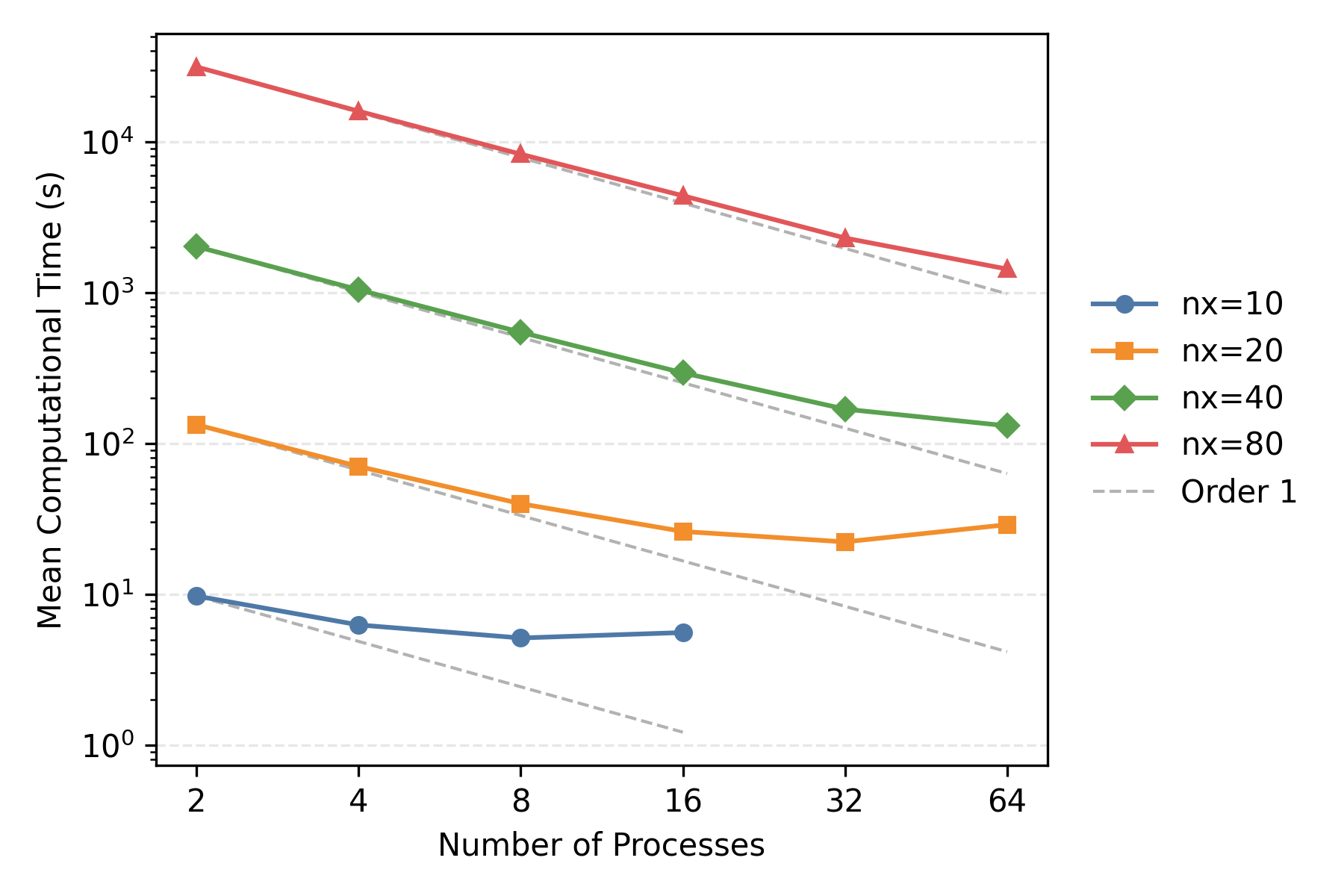}
  \caption{Strong-scaling performance of the domain-level parallelization (MPI) for the \texttt{cylinder2d} case. The plot shows the mean computational time per sample as a function of the number of processes for different resolutions $n_x$.}
  \label{fig:cylinder2d_mpi}
\end{figure}
As shown in Figure~\ref{fig:cylinder2d_mpi}, the performance gain of domain-level parallelization is saturated by communication overhead at different process counts depending on the spatial resolution. 
To ensure efficient resource utilization without incurring excessive parallel overhead, we selected the saturation point for each resolution as the number of processes used in subsequent tests. 
The chosen MPI process counts for each resolution are summarized in Table~\ref{tab:max_mpi_resolution}. 
These configurations are consistently applied in all domain-level simulations and postprocessing runs presented in the following analyses.
\begin{table}[ht!]
  \centering
  \small
  \caption{Selected number of MPI processes at which domain-level parallelization saturates for each spatial resolution $n_x$ in \texttt{cylinder2d}. These configurations are used in all subsequent domain-level simulations and postprocessing.}
  \vspace{0.5em}
  \label{tab:max_mpi_resolution}
    \begin{tabular}{l r r r r}
        \toprule
        \makecell[c]{$n_x$} & \makecell[c]{10} & \makecell[c]{20} & \makecell[c]{40} & \makecell[c]{80} \\
        \midrule
        MPI processes & 2 & 4 & 16 & 64 \\
        \bottomrule
    \end{tabular}
\end{table}

Subsequently, we assessed the overall parallel performance of both sample-level and domain-level decomposition strategies on the \texttt{cylinder2d} case. 
For each resolution, we performed 100 Monte Carlo samples. 
Table~\ref{tab:parallel_speedup} summarizes the average computational time per sample, total simulation time, and the resulting speedup factors relative to the fully sequential execution.
The speedup $S$ achieved by each parallelization strategy is computed as
\begin{equation}
S = \frac{T_{\mathrm{seq}}}{T_{\mathrm{par}}},
\end{equation}
where $T_{\mathrm{seq}}$ denotes the total execution time of the fully sequential baseline and $T_{\mathrm{par}}$ represents the total execution time of the corresponding parallel execution.

\begin{table}[ht!]
  \centering
  \small
  \caption{Comparison of parallel performance for the \texttt{cylinder2d} case under sample-level and domain-level decomposition on a single node. The time per sample represents the average time across all samples.}
  \vspace{0.5em}
  \label{tab:parallel_speedup}
    \begin{tabular}{l l r r r}
        \toprule
        \makecell[c]{$n_x$} & \makecell[c]{Parallelization strategy} & \makecell[c]{Time per sample [s]} & \makecell[c]{Total time [s]} & \makecell[c]{Speedup} \\
        \midrule
        \multirow{3}{*}{10} 
            & Sequential      &   15.92  &   1645.49  & -- \\
            & Sample-level    &   17.24  &   1004.09  & 1.63  \\
            & Domain-level    &    8.99  &   1025.20  & 1.60  \\
        \midrule
        \multirow{3}{*}{20} 
            & Sequential      &  252.21  &  25473.95  & -- \\
            & Sample-level    &  256.54  &   6689.76  & 3.80 \\
            & Domain-level    &   69.55  &   7151.24  & 3.56  \\
        \midrule
        \multirow{3}{*}{40} 
            & Sequential      & 3814.69  & 382457.32  & -- \\
            & Sample-level    & 4041.55  &  28971.26  & 13.20  \\
            & Domain-level    &  291.12  &  29433.72  & 12.99 \\
        \midrule
        \multirow{3}{*}{80} 
            & Sequential      & 61535.14  & 6153514.42 & -- \\
            & Sample-level    & 76213.13  &  152881.41 & 40.25   \\
            & Domain-level    &  1391.92  &  139358.12 & 44.15 \\
        \bottomrule
    \end{tabular}
\end{table}

In addition to the total simulation time, we also analyzed the parallel performance of the postprocessing phase, which aggregates and processes the results from all Monte Carlo samples.
Table~\ref{tab:parallel_postprocessing_cyl} reports the total postprocessing time required for 100 samples at each resolution. 
The results demonstrate that domain-level parallelization significantly reduces postprocessing time compared to the sample-level approach.
Since the postprocessing under sample-level decomposition is executed sequentially—identical to the baseline—we omit it from the domain-level-only comparison.
\begin{table}[ht!]
  \centering
  \small
  \caption{Comparison of parallel postprocessing time for 100 Monte Carlo samples of the \texttt{cylinder2d} case under sample-level and domain-level decomposition on a single node.}
  \vspace{0.5em}
  \label{tab:parallel_postprocessing_cyl}
  \begin{tabular}{l l r}
    \toprule
    \makecell[c]{$n_x$} & \makecell[c]{Parallelization strategy} & \makecell[c]{Total postprocessing time [s]} \\
    \midrule
    \multirow{2}{*}{10} 
        & Sample-level      &  51.37  \\
        & Domain-level      &  24.71  \\
    \midrule
    \multirow{2}{*}{20} 
        & Sample-level      & 202.76  \\
        & Domain-level      &  50.66  \\
    \midrule
    \multirow{2}{*}{40} 
        & Sample-level      & 802.26 \\
        & Domain-level      &  51.32 \\
    \midrule
    \multirow{2}{*}{80} 
        & Sample-level      &  3213.26  \\
        & Domain-level      &  50.27  \\
    \bottomrule
  \end{tabular}
\end{table}

\begin{figure}[ht!]
  \centering
  \includegraphics[width=0.7\textwidth]{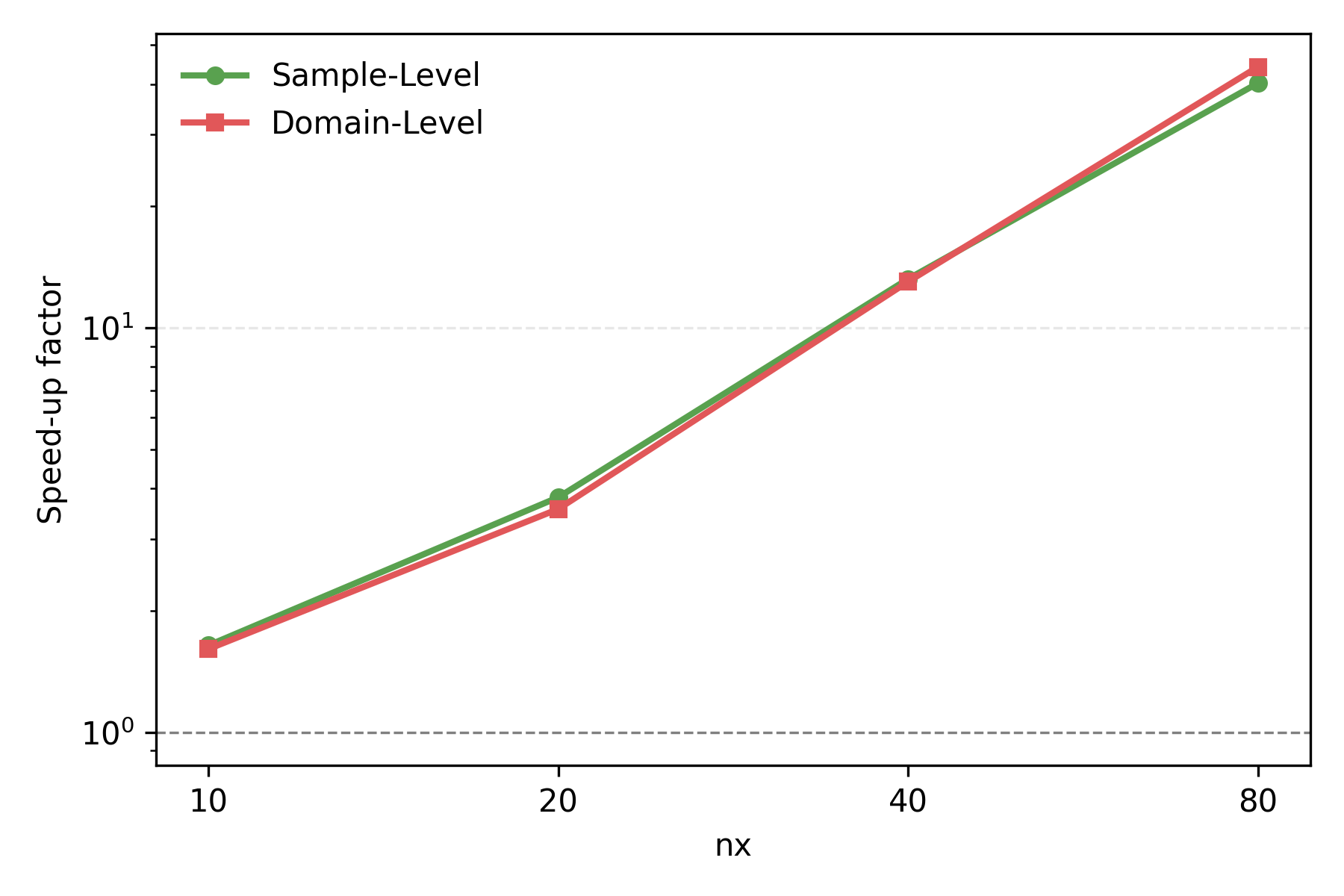}
  \caption{Speedup factors achieved by sample-level and domain-level parallelization strategies for the \texttt{cylinder2d} case at different resolutions $n_x$. The reference baseline for speedup calculation is the sequential execution time.}
  \label{fig:cylinder2d_speedup}
\end{figure}

Figure~\ref{fig:cylinder2d_speedup} presents the overall speedup factors achieved by the two parallelization strategies across different resolutions.

\begin{figure}[ht!]
  \centering
  \includegraphics[width=\textwidth]{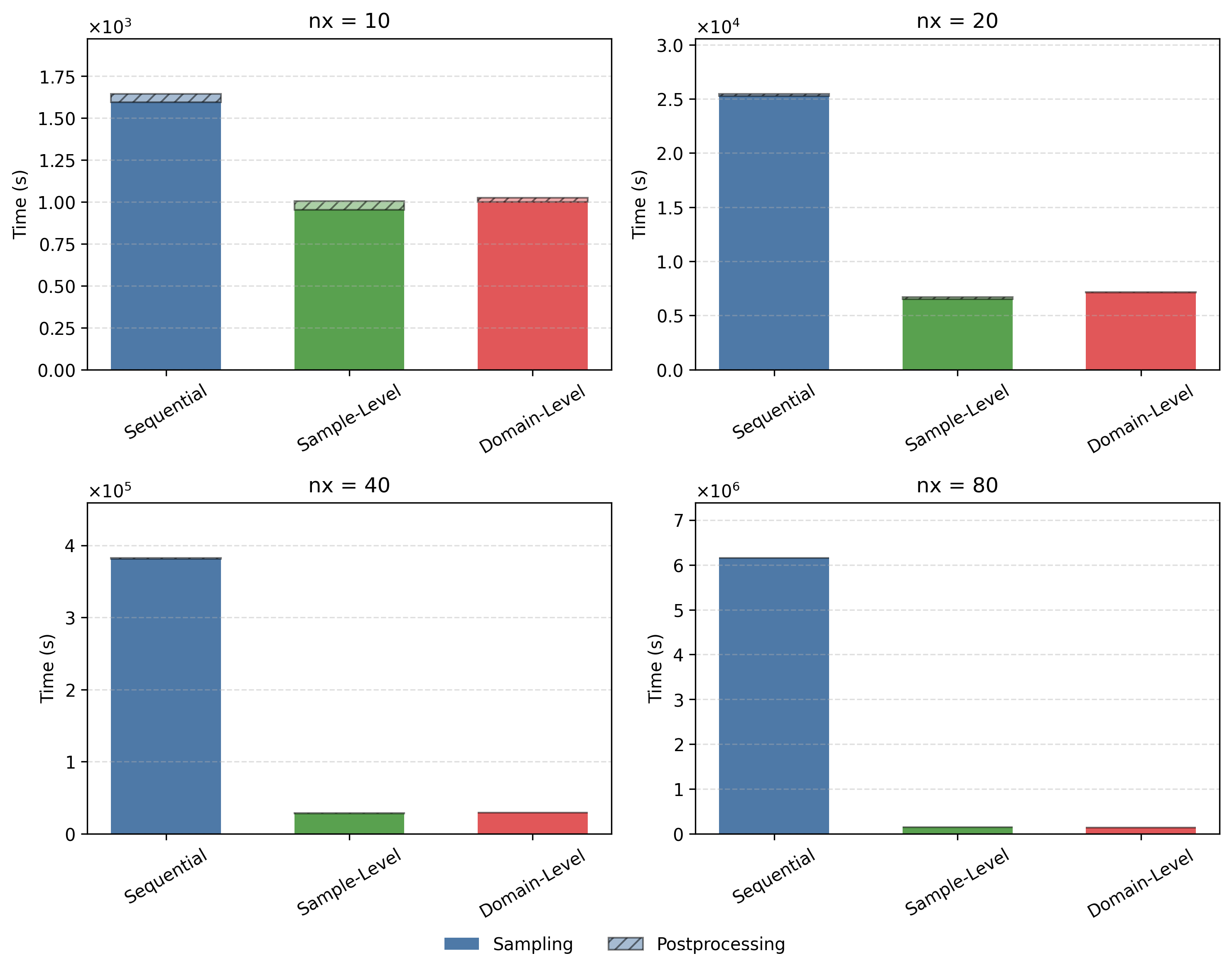}
  \caption{Breakdown of total computational time for 100 Monte Carlo samples of \texttt{cylinder2d} at different resolutions $n_x$, comparing sequential execution, sample-level parallelization, and domain-level parallelization. Each bar shows the sampling time and the postprocessing time.}
  \label{fig:cylinder2d_total_time}
\end{figure}

To visualize the relative contribution of sampling and postprocessing phases under different strategies and resolutions, Figure~\ref{fig:cylinder2d_total_time} shows the breakdown of the total computational time for 100 Monte Carlo samples.

\begin{figure}[ht!]
  \centering
  \includegraphics[width=1.0\textwidth]{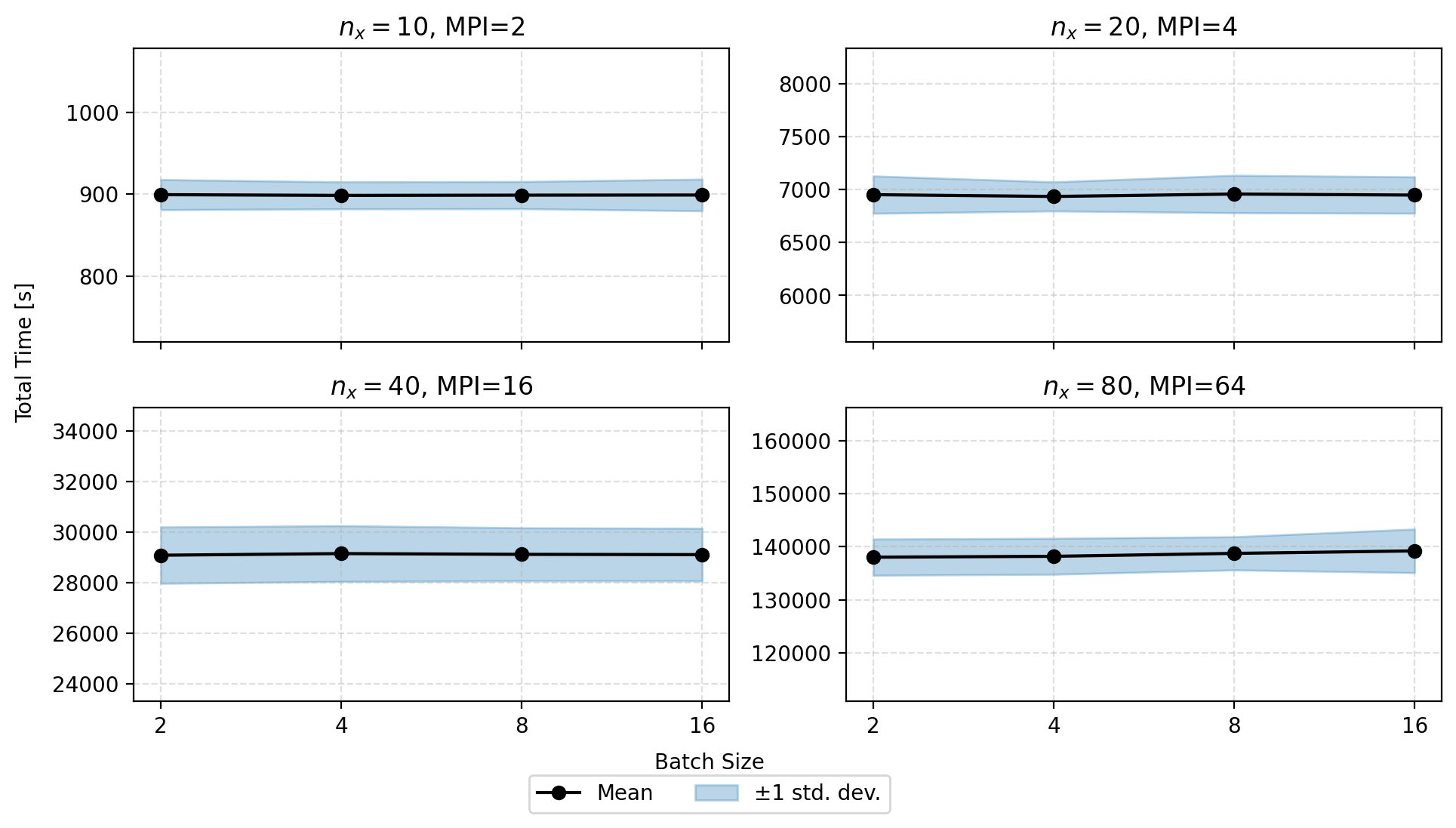}
  \caption{
    Weak scaling performance of sample-level parallelization for the \texttt{cylinder2d} case at several resolutions ($n_x=10,20,40,80$) without postprocessing.
    The black line shows the mean total time per batch size, with the blue shaded area indicating the standard deviation across repeated runs.
    Each MPI process is assigned 100 samples, resulting in batch sizes of 200, 400, 800, and 1600 for 2, 4, 8, and 16 processes, respectively.
    This batch configuration is applied uniformly across all resolutions.    
  }
  \label{fig:cylinder2d_weak_scaling}
\end{figure}

Figure~\ref{fig:cylinder2d_weak_scaling} shows weak scaling results for the \texttt{cylinder2d} case using sample-level parallelization without postprocessing.
Each process is assigned 100 samples, resulting in batch sizes of 200, 400, 800, and 1600 samples for 2, 4, 8, and 16 processes, respectively. 
This batch configuration is kept consistent across all spatial resolutions.
As expected for MCS, the total simulation time remains nearly constant across batch sizes due to the embarrassingly parallel nature of the computation.
Since SC-gPC also employs non-intrusive sampling, this result equally confirms the scalability of SC-based UQ methods.

In summary, the choice between sample-level and domain-level parallelization depends strongly on the problem size and grid resolution. 
For coarse grids, the sample-level decomposition exhibits a clear advantage due to its embarrassingly parallel nature and minimal communication overhead.
However, as the grid resolution increases, the scalability of domain-level (MPI-based) parallelization tends to improve, offering higher speedup for large-scale problems. 
Therefore, selecting the appropriate parallelization strategy requires balancing between available computational resources, problem size, and desired time-to-solution.

\begin{figure}[ht!]
  \centering
  \includegraphics[width=0.96\textwidth]{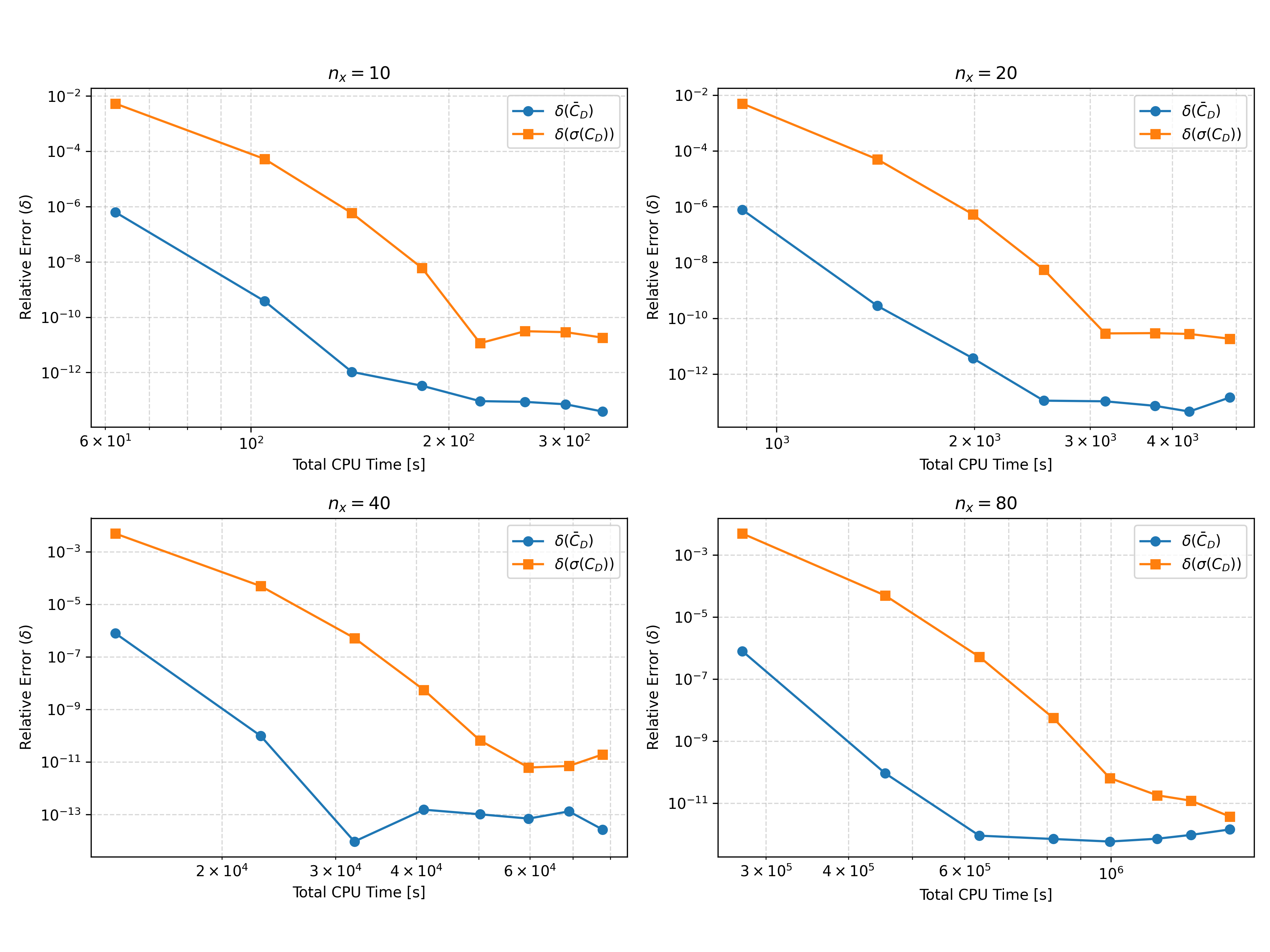}
  \caption{
    Performance of the SC-gPC method for estimating the mean and standard deviation of the drag coefficient \( C_{\mathrm{D}} \) at spatial resolutions \( n_x = 10, 20, 40, 80 \).
    Each data point corresponds to a different polynomial order \( k \in \{1, \dots, 8\} \), and the reference solution is taken from the result at the highest order \( k = 9 \).
    The relative error is plotted against the total CPU time (sampling plus postprocessing).
  }
  \label{fig:scEfficiency}
\end{figure}

To evaluate the computational efficiency of the SC-gPC method, we compare the relative error of the mean and standard deviation of the drag coefficient \( C_{\mathrm{D}} \) with respect to the total CPU time at multiple grid resolutions. 
The results are shown in Figure~\ref{fig:scEfficiency}. 

For the TGV case with four-dimensional uncertainty (\texttt{tgv2d}), a third-order, three-level Smolyak sparse grid quadrature was employed to compute the SC-gPC approximation. 
This configuration required a total of 137 quadrature samples, which provides a reasonable approximation of the target statistics.

\begin{figure}[ht!]
  \centering
  \includegraphics[width=0.7\textwidth]{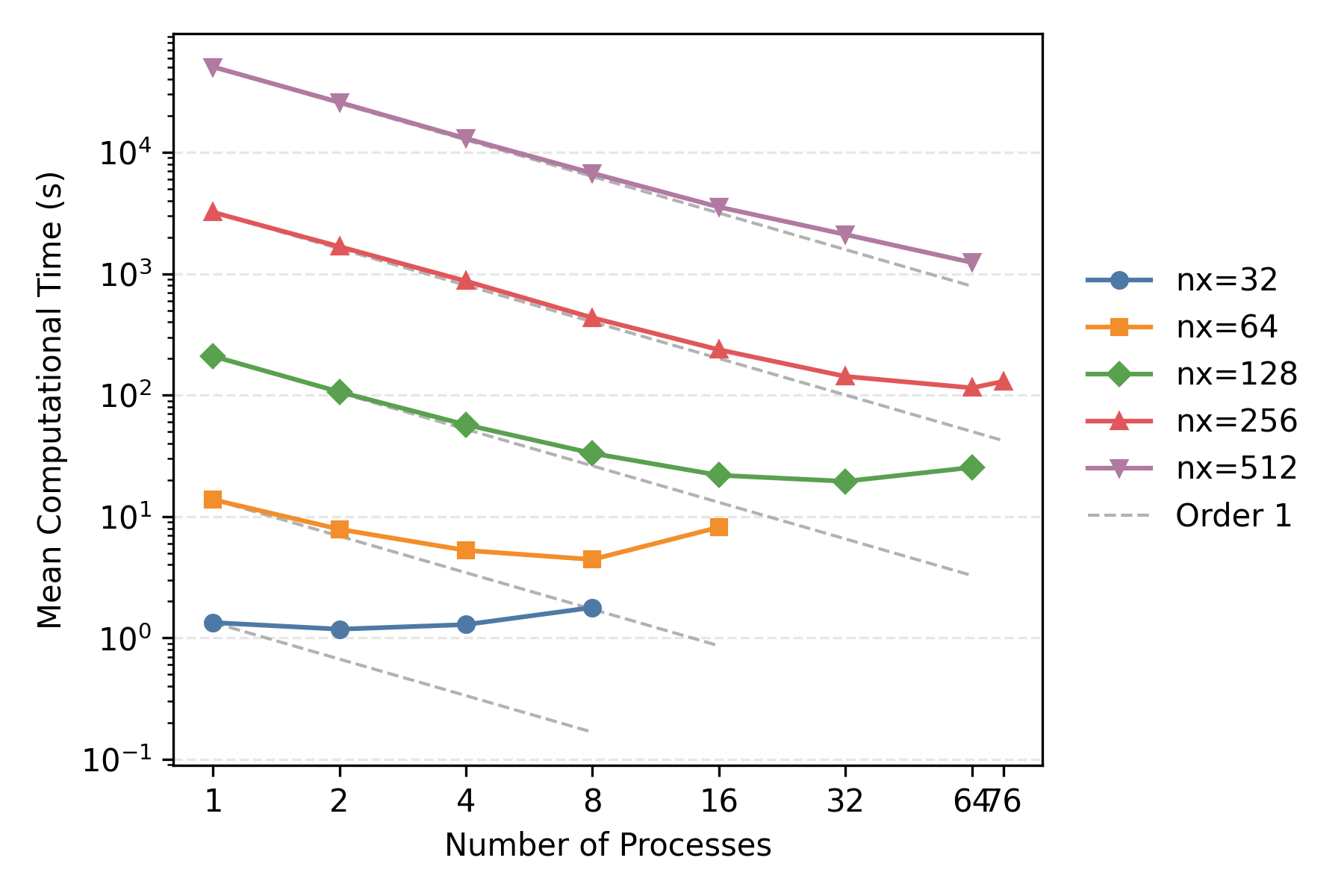}
  \caption{Strong-scaling performance of the domain-level parallelization (MPI) for the \texttt{tgv2d} case. The plot shows the mean computational time per sample as a function of the number of processes for different resolutions $n_x$.}
  \label{fig:tgv2d_mpi}
\end{figure}

To optimize parallel performance, we first evaluated the mean computational time per sample under domain-level MPI decomposition for different spatial resolutions \(n_x \in \{64, 128, 256, 512\}\). 
The goal was to identify the optimal number of MPI processes for each grid size. 
The results, shown in Figure~\ref{fig:tgv2d_mpi}, indicate that process number \(\{2, 8, 32, 64\}\) will get the best parallel performance.

As above, due to the saturation points, the chosen MPI process counts for each resolution are summarized in Table~\ref{tab:max_mpi_resolution_tgv}. 
These configurations are consistently applied in all domain-level simulations and postprocessing runs presented below.
\begin{table}[ht!]
  \centering
  \small
  \caption{Selected number of MPI processes at which domain-level parallelization saturates for each spatial resolution $n_x$ in \texttt{tgv2d}. These configurations are used in all subsequent domain-level simulations and postprocessing.}
  \vspace{0.5em}
  \label{tab:max_mpi_resolution_tgv}
    \begin{tabular}{l r r r r}
        \toprule
        \makecell[c]{$n_x$} & \makecell[c]{64} & \makecell[c]{128} & \makecell[c]{256} & \makecell[c]{512} \\
        \midrule
        MPI Processes & 2 & 4 & 16 & 64 \\
        \bottomrule
    \end{tabular}
\end{table}

Table~\ref{tab:tgv2d_parallel_speedup} summarizes the average time per sample, total simulation time, and speedup relative to the fully sequential baseline for each strategy and resolution. 
As shown, sample-level decomposition exhibits superior performance for fine grids, achieving up to a speedup of 30.78 at \(n_x=512\). 
However, the domain-level strategy demonstrates increasing scalability as the grid size grows. 
For \(n_x=512\), it achieves a speedup of 30.97, closing the gap to sample-level decomposition.
\begin{table}[ht!]
  \centering
  \small
  \caption{Comparison of parallel performance for the tgv2d case under sample-level and domain-level decomposition on a single node. The time per sample represents the average time across all samples.}
  \vspace{0.5em}
  \label{tab:tgv2d_parallel_speedup}
    \begin{tabular}{r l r r r}
        \toprule
        \makecell[c]{$n_x$} & \makecell[c]{Parallelization strategy} & \makecell[c]{Time per sample [s]} & \makecell[c]{Total time [s]} & \makecell[c]{Speedup} \\
        \midrule
        \multirow{3}{*}{64} 
            & Sequential      &  13.79   &  2031.12  & -- \\
            & Sample-level    &  13.31   &  972.00  & 2.08 \\
            & Domain-level    &  7.17  &  1087.00  & 1.86 \\
        \midrule
        \multirow{3}{*}{128} 
            & Sequential      & 209.23  & 28933.79  & -- \\
            & Sample-level    &   210.23 &  7363.00  & 3.92 \\
            & Domain-level    &  56.00 &  7802.00  & 3.70 \\
        \midrule
        \multirow{3}{*}{256} 
            & Sequential      & 3208.57  & 440211.52  & -- \\
            & Sample-level    &  3240.71 & 29153.00  & 15.10 \\
            & Domain-level    & 233.05  & 32232.00 & 13.65 \\
        \midrule
        \multirow{3}{*}{512} 
            & Sequential      & 50440.64  & 5046437.85 & -- \\
            & Sample-level    & 60496.05  & 163902.88 & 30.78 \\
            & Domain-level    & 1180.67  & 162909.53 & 30.97 \\
        \bottomrule
    \end{tabular}
\end{table}

Table~\ref{tab:parallel_postprocessing_tgv} reports the total postprocessing time required for 100 samples at each resolution. 
The results demonstrate a similar behavior as in the \texttt{cylinder2d} case.

\begin{figure}[ht!]
  \centering
  \includegraphics[width=0.6\textwidth]{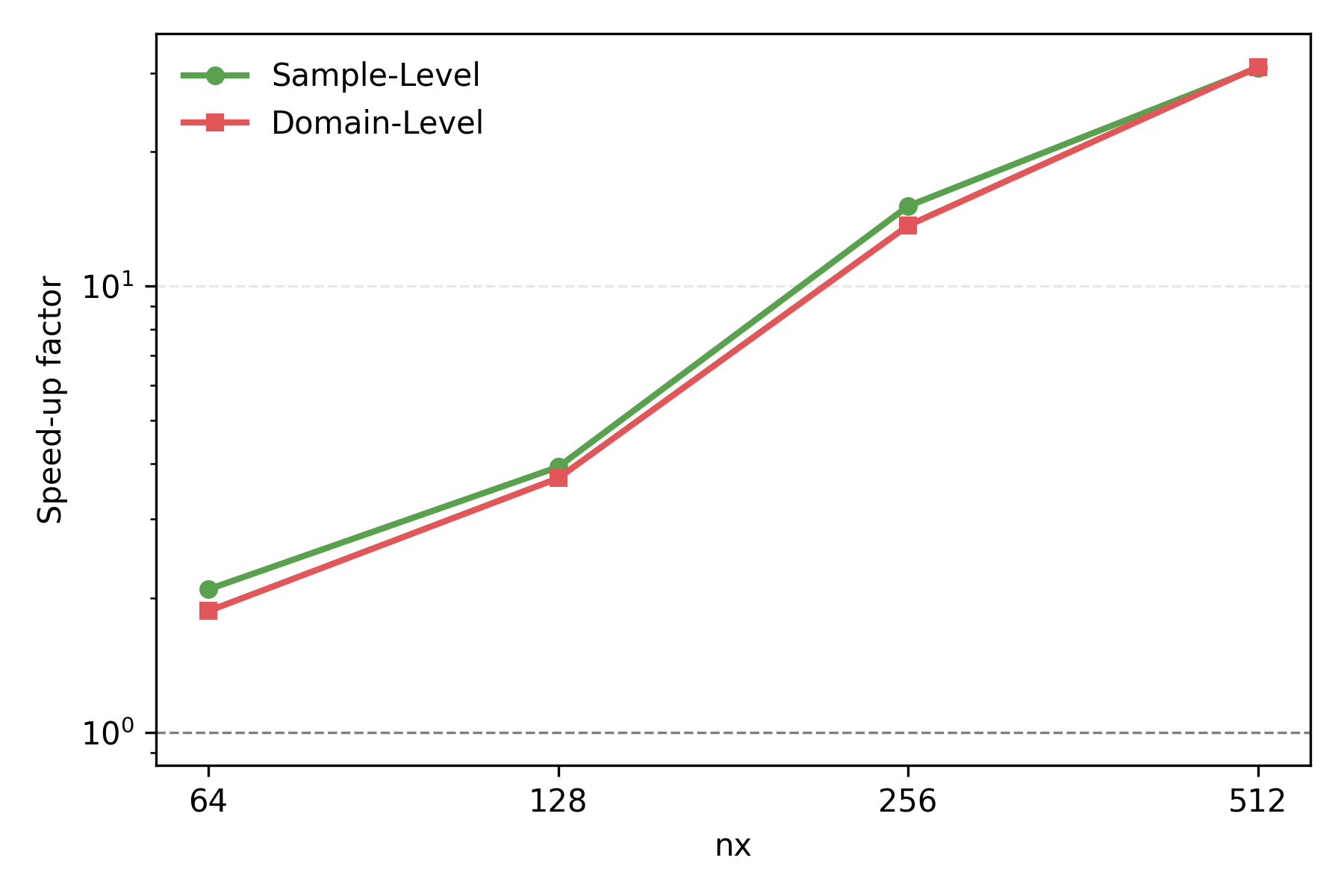}
  \caption{
    Speedup factors achieved by sample-level and domain-level parallelization strategies for the \texttt{tgv2d} case at different resolutions $n_x$.
    The reference baseline for speedup calculation is the sequential execution time.
  }
  \label{fig:tgv2d_speedup}
\end{figure}

Figure~\ref{fig:tgv2d_speedup} presents the overall speedup factors achieved by the two parallelization strategies across different resolutions. 
The sample-level strategy outperforms domain-level parallelization at all resolutions except for \(n_{x}=512\).

Table~\ref{tab:parallel_postprocessing_cyl} reports the total postprocessing time required for 100 samples at each resolution. 
The results demonstrate that domain-level parallelization significantly reduces postprocessing time compared to the sample-level approach.
Since the postprocessing under sample-level decomposition is executed sequentially—identical to the baseline—we omit it from the domain-level-only comparison.

\begin{figure}[ht!]
  \centering
  \includegraphics[width=\textwidth]{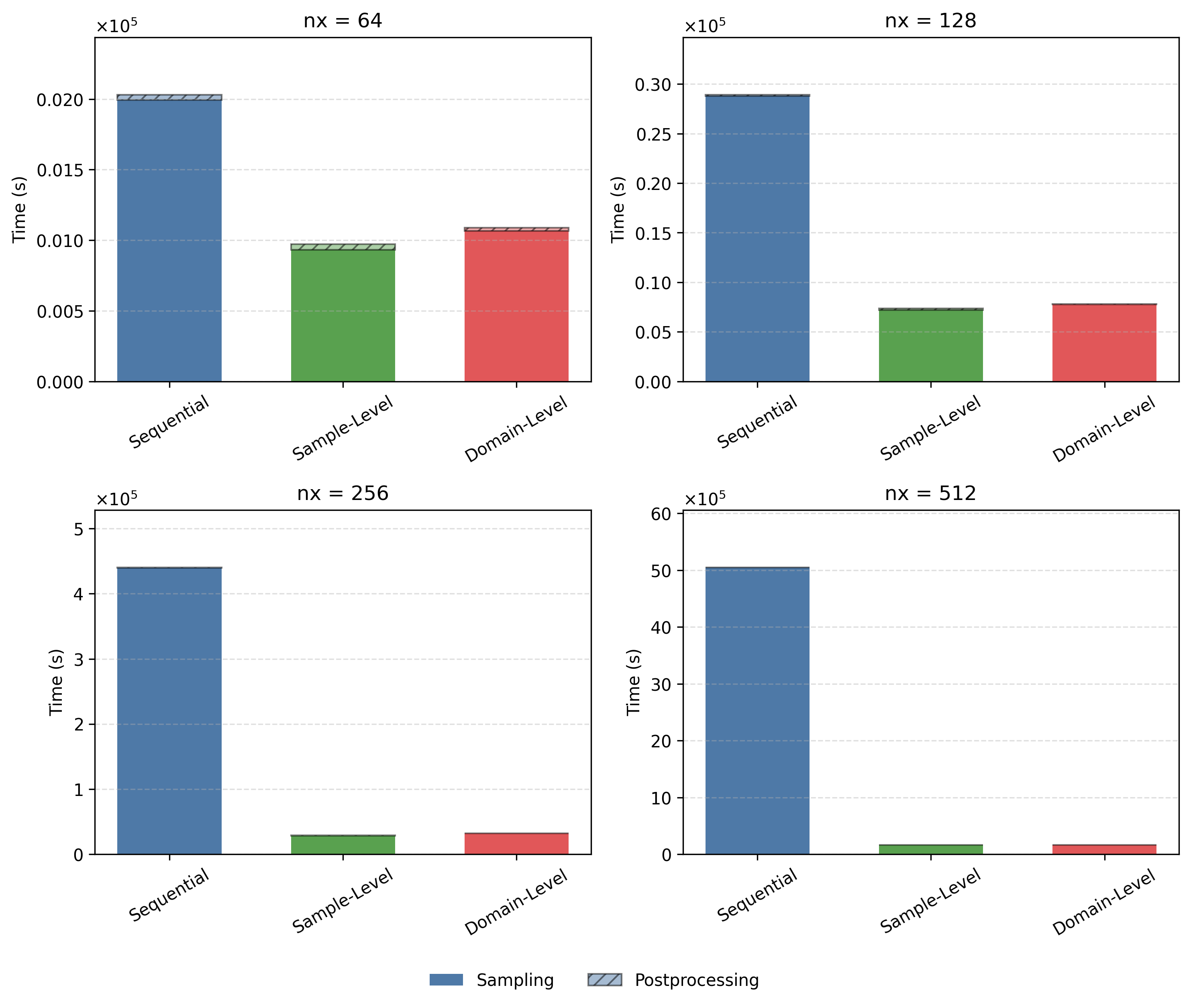}
  \caption{Breakdown of total computational time for 100 Monte Carlo samples of \texttt{tgv2d} at different resolutions $n_x$, comparing sequential execution, sample-level parallelization, and domain-level parallelization. Each bar shows the sampling time and the postprocessing time.}
  \label{fig:tgv2d_total_time}
\end{figure}

To visualize the relative contribution of sampling and postprocessing phases under different strategies and resolutions, Figure~\ref{fig:tgv2d_total_time} shows the breakdown of the total computational time for 100 Monte Carlo samples.

\begin{table}[ht!]
  \centering
  \small
  \caption{Comparison of parallel postprocessing time for third-order, three-level sparse grid samples of the \texttt{tgv2d} case under sample-level and domain-level decomposition on a single node.}
  \vspace{0.5em}
  \label{tab:parallel_postprocessing_tgv}
  \begin{tabular}{l l r}
    \toprule
    \makecell[c]{$n_x$} & \makecell[c]{Parallelization strategy} & \makecell[c]{Total postprocessing time [s]} \\
    \midrule
    \multirow{2}{*}{64} 
        & Sample-level      &  39.139  \\
        & Domain-level      &  20.605  \\
    \midrule
    \multirow{2}{*}{128} 
        & Sample-level      & 150.6  \\
        & Domain-level      &  38.812  \\
    \midrule
    \multirow{2}{*}{256} 
        & Sample-level      & 605.141 \\
        & Domain-level      &  39.119 \\
    \midrule
    \multirow{2}{*}{512} 
        & Sample-level      &  2373.85  \\
        & Domain-level      &  42.530  \\
    \bottomrule
  \end{tabular}
\end{table}

\section{Conclusion}\label{sec:conclusion}

In this paper, we introduce OpenLB-UQ, a modular and fully integrated framework for UQ in incompressible fluid flow simulations using the LBM. 
The framework supports non-intrusive methods such as MCS, QMC, and SC-gPC.
Our numerical experiments on benchmark cases confirm that our MCS implementations are mathematically convergent and that MSC-gPC achieves spectral convergence and offers significant computational advantages. 
The non-intrusive architecture enables seamless integration with existing OpenLB solver modules also for multiphysics simulations, while supporting efficient parallelization through both sample-level and domain-level decomposition.

Although the current implementation supports sparse grids, SC-gPC remains limited by the curse of dimensionality in very high-dimensional settings ($d_Z \gtrsim 5$). 
Future work will aim at extending the framework with broader support for probability distributions and quadrature rules, as well as incorporating multilevel Monte Carlo (MLMC) techniques. 
Overall, \texttt{OpenLB-UQ} provides an efficient and extensible platform for robust UQ in LBM-based simulations.

\section*{Data availability}
The OpenLB-UQ code used in this paper is part of the OpenLB release 1.8.1~\cite{olbRelease18} and is published open source under the GNU General Public License, version 2. 

\section*{Decleration of competing interests}
The authors declare that they have no known competing financial interests or personal relationships that could have appeared to influence the work reported in this paper.

\section*{Author contribution statement}
Conceptualization: MZ, SS; Methodology: MZ, SS; Software: MZ, AK, SI, MJK, SS; Validation: MZ, AK, SS; Formal analysis: MZ, SS; Investigation: MZ, SS; Resources: MJK, MF; Data Curation: MZ, SS; Writing - Original Draft: MZ; Writing - Review \& Editing: AK, SI, MJK, MF, SS; Visualization: MZ, SS; Supervision: MJK, MF, SS; Project administration: SS; Funding acquisition: MJK, MF, SS. 
All authors read and approved the final version of this paper.

\section*{Acknowledgement}
This work has been partially funded from the European Union’s Horizon Europe research and innovation programme under grant agreement No 101138305. 
The authors gratefully acknowledge the computing time provided on the high-performance computer HoreKa by the National High-Performance Computing Center at KIT (NHR@KIT). 
This center is jointly supported by the Federal Ministry of Education and Research and the Ministry of Science, Research and the Arts of Baden-Württemberg, as part of the National High-Performance Computing (NHR) joint funding program (https://www.nhr-verein.de/en/our-partners). 
HoreKa is partly funded by the German Research Foundation (DFG).

\bibliographystyle{elsarticle-num-names} 
\bibliography{ref.bib}

\end{document}